\documentclass[aps,prd,floats,floatfix,onecolumn,nofootinbib,amssymb,amsmath]
{revtex4}
\usepackage{graphicx}
\usepackage{bm}
\usepackage{enumitem}
\graphicspath{ {./} }
\usepackage{hyperref}
\hypersetup{
    colorlinks=true,
    linkcolor=blue,
    filecolor=magenta,      
    urlcolor=cyan,
}
\usepackage{float}
\usepackage{color}
\usepackage[utf8]{inputenc}
\DeclareUnicodeCharacter{2212}{-}
\usepackage[dvipsnames, usenames]{xcolor}
\usepackage[section]{placeins}
\usepackage{ulem}

\newcommand{\half}{{\tfrac{1}{2}}}

\newcommand{\beq}{\begin{equation}}
\newcommand{\eeq}{\end{equation}}
\newcommand{\beqa}{\begin{eqnarray}}
\newcommand{\eeqa}{\end{eqnarray}}

\newcommand{\hateq}{\; \hat{=}\; }

\begin{document}

\title{Numerical Relativity Multimodal Waveforms using Absorbing Boundary Conditions}

\author{Luisa T. Buchman$^{1}$, Matthew D. Duez$^{1}$, Marlo
  Morales$^{1}$, Mark A. Scheel$^{2}$, Tim M. Kostersitz$^{3}$, Andrew M.
  Evans$^{4}$, Keefe Mitman$^{2}$}

\address{$^1$ Department of Physics \& Astronomy, Washington State
  University, Pullman, Washington 99164, USA}
\address{$^2$ TAPIR, Walter Burke Institute for Theoretical Physics,
  MC 350-17, California Institute of Technology, Pasadena, California
  91125, USA}
\address{$^3$ Faculty of Physics, University of Vienna, Boltzmanngasse
  5, A-1090 Vienna, Austria}
\address{$^4$ Department of Physics and Astronomy, UC Irvine, Irvine,
  CA 92697}

\email[Point of Contact: ]{luisa.buchman@wsu.edu}
\date{\today}

\begin{abstract}
  Errors due to imperfect boundary conditions in numerical relativity
  simulations of binary black holes can produce unphysical reflections
  of gravitational waves which compromise the accuracy of waveform
  predictions, especially for subdominant modes.  A system of higher
  order absorbing boundary conditions which greatly reduces this
  problem was introduced in earlier work~\cite{Buchman_2006}.  In this
  paper, we devise two new implementations of this boundary condition
  system in the Spectral Einstein Code ({\tt SpEC}), and test them in
  both linear multipolar gravitational wave and inspiralling mass
  ratio 7:1 binary black hole simulations.  One of our implementations
  in particular is shown to be extremely robust and to produce
  accuracy superior to the standard freezing-$\Psi_0$ boundary
  condition usually used by {\tt SpEC}.
\end{abstract}
\maketitle

\section{Introduction}
\label{intro}

With advanced LIGO-Virgo-Kagra detectors
online~\cite{AdvLIGO_2015,AdvVirgo_2015,KAGRA_2021}, and the
spacecraft LISA in development~\cite{LISA_LRR}, the need to include
accurate non-quadrupolar, subdominant modes of gravitational waves
(GWs) in the waveform models produced by numerical relativity
simulations becomes more and more evident. It has been reported by
many publications (see, for example,~
\cite{LISAConsortiumWaveformWorkingGroup:2023arg, PhysRevD.104.044037,
  PurrerHaster2020, Mehta:2023zlk, Cotesta:2018fcv, LSCandVirgo2020,
  Shaik2020}) that accurate numerical relativity waveforms that
include these higher-order multipoles improve both detection and
parameter estimation for a variety of binary black hole (BBH)
coalescing systems and their remnant properties. Such systems include
unequal mass BBHs, precessing BBHs, binaries whose orbits are inclined
with respect to the observer, those with certain spin alignments, and
those with eccentric orbits. In~\cite{Yang:2023zxk}, it is shown that
correct modeling of subdominant modes enables early warning and
localization of GWs, which is crucial for multi-messenger
astronomy. In~\cite{Pollney:2010hs,Mitman:2020,PhysRevD.80.024002}, the difficulty of
current Cauchy codes to calculate GW non-oscillatory modes and GW
memory effects properly is presented. Studies such
as~\cite{Joshi:2022ocr,MillsFairhurst_2021} show the high increase in
signal-to-noise ratio as a result of including higher-order multipoles
in the waveforms obtained from simulations of eccentric, spinning, BBH
mergers. For the spacecraft LISA, detection of binaries with high
masses and unequal mass ratios is paramount, and also depends on
accurate modeling of subdominant
modes~\cite{Chandra:2022ixv}. Finally, accurate numerical relativity
multimodal waveforms will improve templates obtained using analytic
effective-one-body techniques, phenomenological models, and surrogates
based on reduced order modeling \cite{Islam2021, Colleoni2021}.

To illustrate how subdominant modes can become important for GW
modeling, we take the example of unequal mass BBHs. For these systems,
the amplitudes of the subdominant $(\ell, m)$ modes relative to the
dominant quadrupolar $(2,2)$ mode increase with mass
ratio~\cite{Buchman_2012}. For instance, at the frequency which gives
the highest $(2,2)$ wave amplitude, the ratio of the $(3,3)$ mode
amplitude to the $(2,2)$ mode amplitude increases from $0.14$ to
$0.28$ as the mass ratio increases from 2:1 to 6:1 (see Figure 10
of~\cite{Buchman_2012}). Thus, correct calculation of gravitational
waveforms for higher-order multipoles becomes increasingly critical as
the mass ratio of the BBHs becomes increasingly asymmetric.

Most numerical relativity Cauchy simulations do not evolve the entire
spacetime out to spatial infinity, but instead truncate the domain at
some artificial outer boundary at a far but finite distance from the
source. The treatment of this boundary is a potential source of
error. Evolving the Einstein field equations numerically on a
truncated domain to obtain accurate and unique solutions is a
difficult problem. Numerical and mathematical relativists have been
striving for decades to formulate the Einstein equations so that the
initial boundary value problem is well posed and the outer boundary
conditions during the numerical evolution are constraint-preserving
and, ideally, perfectly absorbing. By perfectly absorbing, we mean
that the outer boundaries are completely transparent to gravitational
waves passing through, including
backscatter. (See~\cite{SarbachTiglio2012} and the references therein
for an excellent review on these topics).  Achieving {\it perfectly}
absorbing boundary conditions in numerical relativity is
unrealistic. However, one can impose {\it approximately} absorbing
boundary conditions and thus significantly reduce the spurious
reflections off the outer boundary which contaminate the numerical
evolution inside the grid.

A hierarchy of boundary conditions for general relativity which is
perfectly absorbing for all multipoles $\ell$ of linearized
gravitational radiation up to a given angular momentum number $L$ was
first introduced in~\cite{Buchman_2006}.  This hierarchy is referred
to as a set of ``higher order boundary conditions'' (HOBCs), where
each boundary condition order is a rung of the hierarchy, numbered by
$L$. The HOBCs presented in~\cite{Buchman_2006} are imposed on the
Newman-Penrose scalar $\Psi_0$, and assume a Minkowski background and
a spherical outer boundary. These HOBCs were subsequently improved
in~\cite{Buchman_2007}, to include first-order corrections for
curvature and backscatter (for quadrupolar radiation), and to be
applicable to fairly general Cauchy foliations and to spherical as
well as non-spherical outer boundary shapes. In this same publication,
James M. Bardeen generalizes these HOBCs for all $\ell \le L$ in the
case of first order curvature corrections (but neglecting
backscatter).

The standard boundary conditions used in the Spectral Einstein Code
{\tt SpEC} ~\cite{kidder_website} for the incoming GW degrees of
freedom at the artificial outer boundary freeze the Newman-Penrose
scalar $\Psi_0$ to its initial value. Referred to simply as
``freezing-$\Psi_0$ boundary conditions'', they are equivalent to the
lowest rung of the hierarchy ($L=1$). It is shown
mathematically~\cite{Buchman_2006} that the freezing-$\Psi_0$ boundary
condition causes some amount of spurious reflection at the outer
boundary, even for the dominant quadrupolar mode ($\ell=2$) of the
GWs, while the HOBCs perfectly absorb linearized radiation for all
modes with angular momentum number $\ell \le L$. Specifically, the
reflection coefficient for the freezing-$\Psi_0$ boundary condition
decays as $(kR_{\rm bdry})^{-4}$ for every $\ell$-mode and for large
$kR_{\rm bdry}$ (where $k$ is the wavenumber of the radiation and
$R_{\rm bdry}$ is the radius of the outer boundary), whereas the
reflection coefficients for the HOBCs are {\it zero} for $\ell \le L$
and further, they decay as $(kR_{\rm bdry})^{-2(L+1)}$ for $\ell > L$
and for large $kR_{\rm bdry}$. (See~\cite{Buchman_2006} for
derivations of these reflection coefficients and their decay rates.)
This means, for example, that the reflection coefficient for the
$\ell = 3$ mode decays as $(kR_{\rm bdry})^{-6}$ when the HOBC with
$L = 2$ is imposed, and the reflection coefficient for $\ell = 4$
decays as $(kR_{\rm bdry})^{-8}$ when the HOBC with $L = 3$ is
imposed. It is apparent from these decay rates that spurious
reflections at a specific outer boundary radius are significantly less
with HOBCs than with the freezing-$\Psi_0$ boundary condition. The
current strategy for reducing reflections in {\tt SpEC} BBH
simulations which utilize the freezing-$\Psi_0$ boundary condition is
to place the outer boundary at moderately far distances (typically
$R_{\rm bdry}\gtrsim 800M$). HOBCs should in principle allow for
smaller $R_{\rm bdry}$'s, thus reducing the computational cost of
simulations.

In \cite{RuizRinneSarbach_2007}, it has been shown that the HOBCs
presented in~\cite{Buchman_2006} are well posed for the second-order
generalized harmonic formulation of the Einstein equations, at least
in the high-frequency limit (see
also~\cite{KreissWinicour_2006,KreissReulaSarbachWinicour2007}). Although
not proved {\it per~se}, it is likely that the HOBCs in the first
order formulation are well posed as well. These HOBCs were
implemented~\cite{Rinne2009} for the first order generalized harmonic
Cauchy formulation~\cite{Lindblom2006} of the Einstein equations in
{\tt SpEC} as algebraic conditions on the Regge-Wheeler-Zerilli (RWZ)
scalars~\cite{ReggeWheeler_1957, Zerilli_1970} in a non-rotating,
single frame grid. Tests were performed with multipolar
waves~\cite{Rinne2009b}, and the expected absorbing features were
demonstrated.  However, previous experience suggests that for BBH
simulations, it is preferable to (i) implement the boundary conditions
on the time derivative of the incoming characteristic fields and (ii)
use a grid coordinate system co-moving with the binary black holes.

In this paper, we enhance the HOBC implementation of~\cite{Rinne2009}
for use with BBH systems.  The HOBCs have been adapted for the
dual-frame infrastructure of {\tt SpEC}~\cite{Scheel2006SolvingEE},
and they have been recast as conditions on the time derivatives of the
incoming physical characteristic modes. Regarding the latter, we have
devised two natural ways of doing this. The first, corresponding to a
generalization of Eq.~(68) of~\cite{Lindblom2006}, uses information
from the HOBC system to set a projection of the Weyl tensor. The
second, corresponding to Equation~(69) of~\cite{Lindblom2006},
directly sets the time derivative of the physical modes on the
boundary to the prediction of the HOBC system. The former is simpler
to implement, reduces exactly to the freezing-$\Psi_0$ condition for
$L=1$, and turns out to be much more robust. Our HOBCs require that a
set of auxiliary ordinary differential equations (ODEs) living on the
outer boundary are integrated forward in time. These ODEs require
initial conditions on the outer boundary.  Accordingly, we devise a
recipe for initializing these ODEs when GWs are present on the
boundary at the initial time (e.g. due to junk radiation).  This
requires imposing the compatibility conditions that express
consistency between the system of ODEs on the boundary and the metric
in the interior volume near the boundary
(see~\cite{SarbachTiglio2012,rauch1974differentiability} for a review
of compatibility conditions in the context of well-posed initial
boundary value problems).

Our new implementations are tested both for GWs on a Minkowski
background and for a mass ratio 7:1 inspiralling binary black hole
system (prior to merger), and compared against the same tests using
the standard $\tt SpEC$ freezing-$\Psi_0$ boundary condition.  We
calculate waveform errors due to boundary conditions by comparing
waveforms extracted on the boundary with analytic solutions (for GWs
on a Minkowski background) or with waveforms extracted at the same
location but in simulations with more distant boundaries (for BBH
simulations). We find the HOBC implementation which sets a projection
of the Weyl tensor (WeylHOBC) to be superior to that which sets the
time derivative of the physical modes (dtHOBC). Furthermore, the
WeylHOBC implementation is found to produce higher waveform accuracy
for quadrupolar and subdominant modes of the GWs than does the
freezing-$\Psi_0$ boundary condition.  In fact, our BBH simulations
with WeylHOBCs give impressively lower errors for the six largest
amplitude modes of GW strain waveforms extrapolated to future null
infinity than those with the freezing-$\Psi_0$ boundary condition. Not
all boundary-related errors are ameliorated by HOBC, however, because
of currently untreated errors in the gauge modes.

Another approach presented in the literature for constructing
absorbing boundary conditions is a technique called Cauchy
characteristic matching (CCM)~\cite{WinicourLRRcce, Bishop1999,
  szilagyi2000ccm, Calabrese_2006, SizhengMa_Aug2023}. This technique
combines Cauchy evolution (which is best suited for simulating the
strong-field regions of spacetime~\cite{1979sgrr83Y}) in the interior
of the domain with characteristic evolution along null hypersurfaces
to future null infinity (where gravitational radiation is defined
unambiguously~\cite{hBmBaM62, rS62, jS89}) in the exterior. Between
the Cauchy and characteristic regions is a timelike interface across
which information can flow in both directions. This timelike interface
acts as an absorbing boundary for the interior Cauchy evolution since
GW data at this boundary is obtained from the
characteristic evolution.  An approach much like our Weyl tensor HOBC
formulation has been recently implemented in~\cite{SizhengMa_Aug2023}
to achieve fully relativistic three-dimensional CCM in the numerical
relativity code SpECTRE~\cite{SpECTRE_zenodo}. This CCM code has
successfully passed several tests in non-trivial numerical relativity
scenarios, but it has not yet been tried in BBH simulations. Further,
it is not clear that the CCM system is well-posed since the
characteristic formulation of the Einstein equations is only weakly
hyperbolic~\cite{PhysRevD.102.064035,PhysRevD.108.104033}.

Our paper is structured as follows.  In Section~\ref{review}, we
review the general HOBC formalism.  In Section~\ref{NeccChanges}, we
present details of our new implementations of the boundary conditions.
Results of multipolar wave tests and binary black hole inspiral
simulations are presented in Section~\ref{results}.  Finally, in
Section~\ref{conclusions} we summarize our findings and suggest future
improvements and applications.

Throughout this paper, Greek indices $\alpha,\beta,\ldots$ are
spacetime indices, lower-case Latin indices $a,b,\ldots$ range over
$t$ and $r$, and upper-case Latin indices $A,B,\ldots$ range over
$\theta$ and $\phi$. Latin indices $i,j,\ldots$ from the middle of the
alphabet are spatial Cartesian coordinates.

\section{Review}
\label{review}

The HOBCs on the RWZ scalars $\Phi^{(\pm)}_{\ell m}$ presented
in~\cite{Rinne2009}, as translated from those on the Newman-Penrose
scalar $\Psi_0$~\cite{Buchman_2006} for the purposes of numerical
relativity, are
\begin{equation}
  \label{e:FlatCL}
  [r^2 (\partial_t + \partial_r)]^{L+1} \Phi^{(\pm)}_{\ell m} \hateq 0,
\end{equation}
where $L$ is the boundary condition order. As developed
in~\cite{Sarbach2001}, the RWZ formalism (originally put forth
in~\cite{ReggeWheeler_1957} and~\cite{Zerilli_1970}) describes
gauge-invariant gravitational perturbations of Schwarzschild
spacetime, although here, we focus on the special case of a flat
rather than Schwarzschild background. The RWZ scalars are denoted by
$\Phi^{(\pm)}_{\ell m}$, where $\Phi^{-}_{\ell m}$ is the odd-parity
Regge-Wheeler scalar, $\Phi^{+}_{\ell m}$ is the even-parity Zerilli
scalar, and the subscripts $\ell,m$ refer to a spherical harmonic
decomposition. They obey the following master equation for flat
spacetime:
\begin{equation}
\label{e:FlatRWZequation}
\left[\partial_t^2-\partial_r^2+\frac{\ell(\ell+1)}{r^2}\right]\Phi^{(\pm)}_{\ell m}=0.
\end{equation}
The HOBCs given in Eq.~(\ref{e:FlatCL}) are perfectly absorbing for
all perturbations with angular momentum numbers $\ell \le L$. Note
that Eq.~(\ref{e:FlatCL}) is the well-known Bayliss-Turkel
conditions~\cite{Bayliss1980} for the scalar wave equation.

As in reference~\cite{Rinne2009}, it is assumed in this paper that the
metric fields can be linearized about flat spacetime close to the
outer boundary, which is taken to be a sphere of constant radius $r$
(which is constant in time as well)\footnote{\label{ConstRbdry}In
  standard {\tt SpEC} BBH simulations, the outer boundary is not kept
  at a constant radius but rather allowed to drift inward at a slow
  velocity. As a consequence of this slow drift, many characteristic
  fields change from zero-speed to outgoing which means that these
  characteristic fields no longer need boundary conditions. As a
  result, reflections that may have been caused by these boundary
  conditions no longer exist. Note that {\it all} the results in this
  paper are for runs with constant outer boundary radius.}. The
spacetime metric $g_{\alpha\beta}$ is written as
\begin{equation}
  g_{\alpha\beta} = \mathring{g}_{\alpha\beta} + \delta g_{\alpha\beta},
\end{equation}
and the background metric $\mathring{g}_{\alpha\beta}$ is
\begin{equation}
\label{e:Background}
\mathring{g} = 
\tilde g_{ab} dx^a dx^b + r^2 \hat g_{\text{\tiny{AB}}} dx^{\text{\tiny{A}}} dx^{\text{\tiny{B}}},
\end{equation}
where $\tilde g=-dt^2+dr^2$ is the standard Minkowski metric on a
2-manifold $\tilde M$ and
$\hat g = d\theta^2 + \sin^2\theta \; d\phi^2$ is the standard metric
on the 2-sphere.  The covariant derivative compatible with the metric
$\tilde g$ ($\hat g$) [$\mathring g$] will be denoted by
$\tilde \nabla$ ($\hat \nabla$) [$\mathring{\nabla}$] and the volume
element by $\tilde \epsilon_{ab}$ ($\hat \epsilon_{\text{\tiny{AB}}}$)
[$\mathring \epsilon_{\alpha \beta \delta \gamma}$].

The RWZ scalars are computed from the spacetime metric. Thus, the
metric perturbations are decomposed with respect to scalar, vector,
and tensor spherical harmonics, using the basis harmonics:
\begin{eqnarray}
  \label{e:YSharmonics}
  &&Y_{\text{\tiny{A}}} \equiv \hat \nabla_{\text{\tiny{A}}} 
     Y,~~~~S_{\text{\tiny{A}}} \equiv \hat
     \epsilon^{\text{\tiny{B}}}{}_{\text{\tiny{A}}} Y_{\text{\tiny{B}}}, 
     \nonumber\\
  &&Y_{\text{\tiny{AB}}} \equiv [\hat \nabla_{({\text{\tiny{A}}}} 
     Y_{{\text{\tiny{B}}})}]^{\tt TT} = 
     \hat \nabla_{({\text{\tiny{A}}}} \hat \nabla_{{\text{\tiny{B}}})}Y 
     + \half \ell(\ell+1) \hat g_{\text{\tiny{AB}}}
     Y,\nonumber\\
  &&S_{\text{\tiny{AB}}} \equiv \hat \nabla_{({\text{\tiny{A}}}} S_{{\text{\tiny{B}}})},
\end{eqnarray}
where $Y$ is the standard scalar spherical harmonic.  The odd and even
parity metric perturbations are treated separately (see
~\cite{Rinne2009} and~\cite{Sarbach2001} for details).  Following the
gauge-invariant procedure of~\cite{Sarbach2001}, the odd parity
perturbations are
\begin{equation}
  \label{e:OddDecomp}
  \delta g_{\text{\tiny{A}} b} = h_b S_{\text{\tiny{A}}},~~~~
  \delta g_{\text{\tiny{AB}}} = 2 k S_{\text{\tiny{AB}}},~~~~
  \delta g_{ab} = h_{ab}=0
\end{equation}
where $h_a$, $k$ and $h_{ab}$ are metric amplitudes, and the even
parity perturbations are
\begin{equation}
  \label{e:EvenDecomp}
  \delta g_{\text{\tiny{A}} b} = Q_b Y_{\text{\tiny{A}}},~~~~
  \delta g_{\text{\tiny{AB}}} =
  r^2(K\hat{g}_{\text{\tiny{AB}}}Y+GY_{\text{\tiny{AB}}}),~~~~
  \delta g_{ab} = H_{ab}Y
\end{equation}
where again $Q_a$, $K$, $G$, and $H_{ab}$ are metric amplitudes. In
terms of these metric amplitudes and for $\ell \ge 2$, the odd parity
Regge-Wheeler scalar is
\begin{equation}
\label{e:Regge-Wheeler}
\Phi^{-}_{\ell m}=r\left(\dot{h}_r-h'_t+2h_t/r\right)/\Lambda,
\end{equation}
and the even parity Zerilli scalar is
\begin{equation}
  \label{e:Zerilli}
  \Phi^{+}_{\ell m}=\left[\Lambda^{-1}\left(2r H_{rr}-2r^2 K'-r^2 \ell(\ell+1)
      G'-4Q_r+2 r^2 G'\right)+r K+r\ell (\ell+1) G/2 -
    2 Q_r+r^2 G'\right]/\ell (\ell+1),
\end{equation}
where a dot denotes partial differentiation with respect to $t$, a
prime denotes partial differentiation with respect to $r$, and
$\Lambda \equiv (\ell-1)(\ell+2)$. The sign difference between
Eq.~(\ref{e:Zerilli}) above and Eq. (29) of~\cite{Rinne2009} allows us
to match our sign conventions with those that have become standard in
the numerical relativity community, as detailed in Appendix C
of~\cite{Boyle:2019kee}, pages 43 and 44.

As was shown in~\cite{Rinne2009}, the HOBCs given in
Eq.~(\ref{e:FlatCL}) can be implemented numerically by introducing a
set of auxiliary variables which are defined only at the
boundary. These are
\begin{equation}
  \label{e:AuxVars}
  \text{w}_ {k\ell m}^{(\pm)}
  \equiv
  r^{-(2 k+1)}[r^2 (\partial_t + \partial_r)]^k \Phi^{(\pm)}_{\ell m}.
\end{equation}
It was further shown in~\cite{Rinne2009} that the these auxiliary
variables obey a system of ODEs on the boundary, namely
\begin{equation}
  \label{e:AuxODEs}
  \partial_t  \text{w}_{k\ell m}^{(\pm)} =
  \frac{1}{2 r^2} [-\ell(\ell+1) + k(k-1)] \text{w}_{(k-1)\ell m}^{(\pm)} 
  + \half \text{w}_{(k+1)\ell m}^{(\pm)} -  \frac{k}{r} \text{w}_{k\ell m}^{(\pm)}.
\end{equation}
This system of ODEs is closed by
\begin{equation}
  \label{e:wL+1BC}
  \text{w}_{(L+1)\ell m}^{(\pm)} \hateq 0,
\end{equation}
which is equivalent to the boundary condition given by
Eq.~(\ref{e:FlatCL}). Eq.~(\ref{e:AuxODEs}) is integrated on the
boundary for $1 \leqslant k \leqslant L$, using
Eq.~(\ref{e:wL+1BC}) and
$\text{w}_{0\ell m}^{(\pm)} = \Phi^{(\pm)}_{\ell m}/r$.

Thus, simulations with HOBCs can be carried out by evolving two
coupled systems: the partial differential equations for the metric in
the interior (the {\it interior system}) and the set of ODEs for
$\text{w}_{k\ell m}^{(\pm)}$ on the boundary (the {\it auxiliary
  system}).

\section{Compatibility Conditions and New Formulations}
\label{NeccChanges}

The absorbing HOBCs in~\cite{Rinne2009} were implemented in a single,
non-rotating frame and tested for multipolar waves~\cite{Rinne2009b}
originating in the interior of the computational domain. The research
presented in this paper extends what was done in~\cite{Rinne2009} to
be applicable to BBH simulations in several ways. First, the HOBC
implementation has been updated for the rotating, dual-frame
infrastructure of {\tt SpEC} used for BBH simulations. Second, we
studied cases where the values of the auxiliary variables on the
boundary were nonzero at $t=0$, mimicking the initial data of BBHs
which contain junk radiation. We found that in these scenarios, the
auxiliary variables had to be initialized properly in order to satisfy
compatibility conditions (see section~\ref{InitAuxVars}). Finally, the
algebraic formulation presented in~\cite{Rinne2009} has been modified
to the form of Eq.~(68) or Eq.~(69) of~\cite{Lindblom2006}. Note that
Eq.~(68) of~\cite{Lindblom2006} is currently used for the
freezing-$\Psi_0$ boundary condition in {\tt SpEC}. We have named the
HOBC formulation derived using Eq.~(68) ``WeylHOBC'', and that derived
using Eq.~(69) ``dtHOBC''. These are discussed in sections~\ref{Weyl}
and~\ref{TimeDeriv} below.

\subsection{Compatibility Conditions}
\label{InitAuxVars}

HOBC simulations evolve two coupled systems, {\it interior} and {\it
  auxiliary}, both of which provide information about the RWZ scalars
$\Phi^{(\pm)}_{\ell m}$.  For the two systems to be consistent, the
values of $\text{w}_{k\ell m}^{(\pm)}$ evolved on the boundary should
agree with the corresponding derivatives of the
$\Phi^{(\pm)}_{\ell m}$ in the interior grid evaluated at the
boundary.  Thus, one can consider Eq.~(\ref{e:AuxVars}) as a
constraint establishing the compatibility of the two systems, a
constraint known as a compatibility condition (see Sec. 5
of~\cite{SarbachTiglio2012}).  We must initialize
$\text{w}_{k\ell m}^{(\pm)}$ correctly for the compatibility condition
to be satisfied at $t=0$.

We introduce the null coordinates $v = t + r$ and $u = t - r$ to
parameterize the numerical solution for the differential operator in
Eq.~(\ref{e:AuxVars}).  In these coordinates, the differential
operator $r^{2}(\partial_{t} + \partial_{r})$ represents a directional
derivative along a path of constant $u$.  For a general choice of
parameter $\lambda(v)$ labeling points on this path, the
corresponding tangent vector will be given by
\begin{equation}
  \label{eq:ddA}
  \left. \frac{\partial}{\partial \lambda} \right|_{u} =
    \left. \frac{\partial r}{\partial \lambda}\right|_{u} (\partial_{r} + \partial_{t}).
\end{equation}
For appropriate choice of $\lambda(r)$, namely that for which
$\left.\partial r/\partial\lambda\right|_u = r^2$,
the derivative $d/d\lambda$ will be the
differential operator $r^{2}(\partial_{t} + \partial_{r})$ in
Eq.(\ref{e:AuxVars}).  Solving for $\lambda(r)$,
\begin{equation}
  \label{eq:A_r}
  \lambda = - \frac{1}{r}.
\end{equation}
We then utilize Eq.~(\ref{eq:ddA}) on our boundary condition formulation
to solve for the initialization of the auxiliary variables.
\begin{equation}
  \label{eq:AuxVars_dA}
  \text{w}_ {k\ell m}^{(\pm)} \equiv r^{-(2 k+1)}
  \left(\frac{\partial}{\partial \lambda}\right)^{k} \Phi^{(\pm)}_{\ell m}.
\end{equation}

Our goal now is to gather $\Phi^{(\pm)}_{\ell m}(\lambda)$ for
particular events along a line of constant $u$. This involves running
a short-time Cauchy evolution on a larger grid extending beyond the
desired boundary location. The larger grid for the short run is chosen
to have an outer boundary sufficiently distant from our desired outer
boundary so as to remain out of causal contact throughout the short
run.  Since the outer boundary of the short run does not matter, it
can use the $L=1$ (freezing-$\Psi_0$) boundary condition which does
not depend on initialization of $\text{w}_{k\ell m}^{(\pm)}$.
Subsequently, we measure $\Phi^{(\pm)}_{\ell m}$ at several values of
$\lambda$ from the short run, and perform a polynomial fit to
$\Phi^{(\pm)}_{\ell m}(\lambda)$:
\begin{equation}
  \label{eq:Phi_A}
  \Phi^{(\pm)}_{\ell m}(\lambda) = \sigma_{0} + \sigma_{1}\lambda + \sigma_{2}\lambda^{2} + 
  \sigma_{3}\lambda^{3}.
\end{equation}
The Python \texttt{scipy.optimize.curvefit} function is employed to determine
$\sigma_{i}$ ($i=$0,1,2,3) for achieving a least squares best
fit. Subsequently, we evaluate $\text{w}_{k\ell m}^{(\pm)}$ using
Eq.(\ref{eq:AuxVars_dA}) in conjunction with Eq.(\ref{eq:Phi_A}).

Recall that our HOBC (Eq.~(\ref{e:FlatCL})) is equivalent to
$\text{w}_{(L+1)\ell m}^{(\pm)}=0$ (Eq.~(\ref{e:wL+1BC})). However,
whereas $\text{w}_{(L+1)\ell m}^{(\pm)}=0$ is enforced in the
auxiliary system, the corresponding condition on the derivatives of
$\Phi^{(\pm)}_{\ell m}$ is not imposed in the interior evolution.  So
the enforcement of Eq.~(\ref{e:wL+1BC}) at each timestep and for each
$\ell$, $m$, and parity is at least somewhat inconsistent with with
the compatibility condition given in~Eq.~(\ref{e:AuxVars}).  One might
impose the compatibility condition for order $(L+1)$ by setting
$\text{w}_{(L+1)\ell m}^{(\pm)}=g(t)$, where $g(t=0)\equiv g_0$ is
determined by the volume data metric evaluated on the boundary for the
initial time. The time dependence must then be specified
explicitly. Setting $g$ to a constant is undesirable, since then the
wave passing through at $t=0$ would leave a permanent imprint on the
boundary. A better alternative might be to damp this initial wave
exponentially by employing something of the form
$g(t)=g_0 e^{-t/\tau}$.  This still will not guarantee satisfaction of
the compatibility condition for $\text{w}_{(L+1)\ell m}^{(\pm)}$ at
later times. It can be argued that if $\text{w}_{(L+1)\ell m}^{(\pm)}$
is dynamically significant, it is advisable to evolve the HOBC
auxiliary system to an order of at least $(L+1)$.  In all cases we
have investigated, we have found that the magnitude of
$\text{w}_{k\ell m}^{(\pm)}$ decreases rapidly with $k$, usually
approaching machine double precision by $k=4$, and furthermore, that
the computational cost of increasing $L$ to this value (or higher) is
negligible.  Consequently, we always set $g_0=0$.

\subsection{New Higher Order Boundary Condition Formulations}
\label{BcFormulations}

In the first order generalized harmonic formulation of the Einstein
equations~\cite{Lindblom2006} used by {\tt SpEC}, there are three
types of characteristic fields which require outer boundary
conditions: constraint, physical, and gauge. As presented
in~\cite{Lindblom2006}, constraint-preserving, freezing-$\Psi_0$, and
constant-in-time gauge boundary conditions are the current standard in
{\tt SpEC}.  In this paper, we focus on improving the boundary
conditions for the characteristic fields representing physical
gravitational waves. We start by outlining the mathematical
description of these physical modes incoming at the outer
boundary. They are given in Eq.~(33) of~\cite{Lindblom2006} as
\begin{equation}
\label{IncomingFieldsAsPerLindblomEtAl}
  u^{1-}_{\alpha\beta} = \Pi_{\alpha\beta} - n^i \Phi_{i\alpha\beta} -
  \gamma_2 g_{\alpha\beta},
\end{equation}
where 
\begin{eqnarray} 
  \Pi_{\alpha\beta} \equiv -t^\gamma \partial_\gamma
                             g_{\alpha\beta}~~~~{\rm and}~~~~
  \Phi_{i\alpha\beta} \equiv \partial_i g_{\alpha\beta}.
\label{e:MetricDerivs}
\end{eqnarray}
In these equations, $\gamma_2$ is a parameter arising from the
addition of constraint-damping terms to the evolution equations,
$n^{\alpha}$ is the outward-pointing unit spacelike normal to the
boundary on the $t=$ constant slices, and $t^{\alpha}$ is the future
directed unit timelike normal to the $t=$ constant surfaces
($t^{\alpha}n_{\alpha}=0$).

In order to cast Eq.~(\ref{IncomingFieldsAsPerLindblomEtAl}) in terms
of covariant derivatives with respect to the background metric instead
of partial derivatives (which depend on the coordinate system), we
introduce the outgoing and incoming null vectors as
$\ell^\alpha \equiv (t^\alpha + n^\alpha)$ and
$k^\alpha \equiv (t^\alpha - n^\alpha)$, respectively, and let
$\mathring{\nabla}_{\ell}\equiv \ell^{\alpha}
\mathring{\nabla}_{\alpha}$ denote covariant differentiation along
$\ell^\alpha$. A simple partial derivative operator
$(\partial_t+\partial_r)$ will be denoted by $D$. The incoming
characteristic fields at the boundary can then be expressed as
\begin{equation}
  u^{1-}_{\alpha \beta} = -(\mathring{\nabla}_{\ell} + \gamma_2) g_{\alpha \beta}.
   \label{e:U1-}
\end{equation}
Note that the introduction of a flat-space covariant derivative in
Eq.~(\ref{e:U1-}) simplifies the transformation to spherical-polar
coordinates.
  
 Given that the 2-metric intrinsic to the boundary is
 $P_{\alpha \beta} \equiv g_{\alpha \beta} + t_{\alpha} t_{\beta} -
 n_{\alpha} n_{\beta}$, the projection operator for the boundary
 conditions on the two physical degrees of freedom is
 $P^\mathrm{P}_{\alpha\beta}{}^{\gamma\delta} = P_\alpha{}^\gamma
 P_\beta{}^\delta - \half P_{\alpha\beta} P^{\gamma\delta}$, where
 $P^\mathrm{P}_{\alpha\beta}{}^{\gamma\delta}$ constructs the
 transverse-traceless version of a symmetric spacetime tensor. Given
 that only the perturbation of the metric is involved, the projection
 is onto the background geometry which in this case is the
 2-sphere. The resulting non-zero components of the physical boundary
 conditions are then the angular components:
\begin{equation}
\label{e:Algebraic}
u^{1-}_{\text{\tiny{AB}}} ~
\hat{=}~ P^\mathrm{P}_{\text{\tiny{AB}}}{}^{\gamma\delta} u^{1-}_{\gamma
  \delta} =
-r^2 (\partial_t + \partial_r)
(r^{-2} \delta g^{\tt TT}_{\text{\tiny{AB}}})-\gamma_2 \; \delta
g^{\tt TT}_{\text{\tiny{AB}}},
\end{equation}
where $\tt TT$ denotes the trace-free part with respect to the metric
$\hat{g}$ on the 2-sphere and $\hat{=}$ indicates ``at the
boundary''.

In~\cite{Rinne2009}, HOBCs are imposed using the auxiliary system of
$\text{w}_{k\ell m}^{(\pm)}$'s to specify the fields
$u_{\text{\tiny{AB}}}^{1-}$ on the boundary, according to Eqs.~(46)
and (50) of that paper.  This was shown in~\cite{Rinne2009} to be
successful when the multipolar wave initial data was zero on the
boundary. However, in the general binary black hole case, there is
non-zero GW data on the boundary initially. As
discussed in~\cite{Kidder_2005} and~\cite{Lindblom2006},
discontinuities produced with an algebraic boundary condition are
avoided by casting the boundary condition in a time-derivative form as
per Bj{\o}rhus~\cite{Bjorhus1995}. Hence the boundary conditions which
we present in the next two sections are represented by
\begin{equation}
\label{e:dtu1-}
  d_t u^{1-}_{\text{\tiny{AB}}} \hat{=}~
  P^\mathrm{P}_{\text{\tiny{AB}}}{}^{\gamma\delta} \partial_t u^{1-}_{\gamma\delta} .
\end{equation}
Reference~\cite{Lindblom2006} presents two ways to implement such time
derivative physical boundary conditions in Eqs.~(68) and~(69),
reproduced here for clarity. Utilizing our notation, Eq.~(68)
of~\cite{Lindblom2006} is
\begin{equation}
\label{e:Eq-68}
d_t u^{1-}_{\text{\tiny{AB}}}\hat{=}~P^\mathrm{P}_{\text{\tiny{AB}}}{}^{\gamma \delta} 
[D_t u^{1-}_{\gamma \delta} + v(\omega_{\gamma \delta} -
\omega_{\gamma \delta}|_{\text{\tiny{BC}}}-\gamma_2 n^ic^3_{i \gamma \delta})],
\end{equation}
where the quantity $D_t u$ is the right hand side of the first order
generalized harmonic evolution system, $v$ is the characteristic speed
at which the characteristic field $u^{1-}_{\alpha\beta}$ enters the
boundary, $\omega_{\gamma \delta}$ is the projection of the Weyl
tensor, $\omega_{\gamma \delta}|_{\text{\tiny{BC}}}$ is the target
value to which $\omega_{\gamma \delta}$ is to be fixed at the
boundary, and $n^ic^3_{i \gamma \delta}$ are incoming constraint
fields (see~\cite{Lindblom2006} for more details). Eq.~(69)
of~\cite{Lindblom2006} is
\begin{equation}
\label{e:Eq-69}
d_t u^{1-}_{\text{\tiny{AB}}}\hat{=}~P^\mathrm{P}_{\text{\tiny{AB}}}{}^{\gamma \delta} 
\dot{h}_{\gamma \delta}(t,r),
\end{equation}
where $\dot{h}_{\gamma \delta}(t,r)$ is a pre-determined waveform
($\dot{x}\equiv \partial_t{x}$).  The precise form of these conditions
and their interface with the auxiliary system will be described in the
two sections below.  The WeylHOBC formulation uses the auxiliary
system to set a projection of the Weyl tensor and is equivalent to
using the time derivative of $u^{1-}$ to drive the radial derivative
of $u^{1-}$ at the boundary to a desired value. The dtHOBC formulation
directly sets the time derivative of the incoming physical field to
its value in the auxiliary system.

\subsubsection{WeylHOBC Formulation}
\label{Weyl}

The incoming wave projection of the Weyl tensor contains the
information needed to construct the physical boundary conditions, as
proposed in~\cite{BardeenBuchman_2002}. These ideas as well as those
in~\cite{ReulaSarbach_2005} were adapted for the 3D Einstein equations
in~\cite{Kidder_2005} and further developed for the first order
generalized harmonic system in~\cite{Lindblom2006}.

The goal of this section is to calculate the quantity
$\omega_{\gamma \delta}|_{\text{\tiny{BC}}}$ of
Eq.~(\ref{e:Eq-68}). To begin, recall
\begin{equation}
\label{e:prop-weyl}
\omega_{\alpha \beta} =
P^\mathrm{P}_{\alpha \beta}{}^{\varepsilon \zeta} ~C_{\varepsilon \mu
  \zeta \nu}\; \ell^\mu \ell^\nu,
\end{equation}
where $C_{\varepsilon \mu \zeta \nu}$ is the Weyl tensor. Since a
vacuum background is assumed and since the projection operator selects
only the angular components, Eq.~(\ref{e:prop-weyl}) becomes
\begin{eqnarray}
  \omega_{\text{\tiny{AB}}} 
  &=&
      P^\mathrm{P}_{\text{\tiny{AB}}}{}^{\varepsilon \zeta} ~R_{\varepsilon \mu
      \zeta \nu}\; \ell^\mu \ell^\nu \\
  &=& \left(R_{\text{\tiny{A}} \mu \text{\tiny{B}} \nu}
      -\half R^{\text{\tiny{C}}}_{\;\mu {\text{\tiny{C}}}
      \nu}P_{\text{\tiny{AB}}}\right) \ell^\mu \ell^\nu, \label{e:Riemann}
\end{eqnarray}
where $R_{\varepsilon \mu \zeta \nu}$ is the Riemann tensor and
$P_{\text{\tiny{AB}}}=r^2 \hat{g}_{\text{\tiny{AB}}}$. In terms of
metric perturbations with respect to the flat spacetime metric in
Eq.~(\ref{e:Background}), Eq.~(\ref{e:Riemann}) becomes
\begin{equation}
\label{e:ang-prop-weyl}
\begin{split}
  \omega_{\text{\tiny{AB}}} = \half \; \ell^\nu \ell^\mu \; \bigl[
  \hat{\nabla}_{\text{\tiny{B}}} \mathring{\nabla}_{\nu} \;\delta
  g_{\text{\tiny{A}} \mu} +\mathring{\nabla}_{\mu}
  \hat{\nabla}_{\text{\tiny{A}}} \;\delta g_{\text{\tiny{B}} \nu}
  -\hat{\nabla}_{\text{\tiny{B}}}\hat{\nabla}_{\text{\tiny{A}}}
  \;\delta g_{\mu \nu}
  -\mathring{\nabla}_{\nu}\mathring{\nabla}_{\mu}\;\delta
  g_{\text{\tiny{AB}}} \\
  -\half \bigl(\hat{\nabla}_{\text{\tiny{C}}} \mathring{\nabla}_{\nu}
  \;\delta g^{\text{\tiny{C}}}_{~ \mu}+
  \mathring{\nabla}_{\mu}\hat{\nabla}_{\text{\tiny{C}}}\;\delta
  g^{\text{\tiny{C}}}_{~ \nu}-\hat{\nabla}_{\text{\tiny{C}}}
  \hat{\nabla}^{\text{\tiny{C}}} \;\delta g_{\mu
    \nu}-\mathring{\nabla}_{\mu} \mathring{\nabla}_{\nu} \;\delta
  g^{\text{\tiny{C}}}_{~{\text{\tiny{C}}}}\bigr) r^2
  \hat{g}_{\text{\tiny{AB}}} \bigr].
\end{split}
\end{equation}
After a lengthy calculation, Eq.~(\ref{e:ang-prop-weyl}) can be
simplified for general parity. Recalling
$\mathring{\nabla}_{\ell}\equiv \ell^{\varepsilon}
\mathring{\nabla}_{\varepsilon}$ (note that $\ell$ is not an index)
and $D=(\partial_t+\partial_r)$,
the result is
\begin{equation}
  \label{e:first-line-simp}
\begin{split}
  \omega_{\text{\tiny{AB}}} 
  = 
  \half  \bigl[\hat{\nabla}_{\text{\tiny{B}}}\mathring{\nabla}_{\ell}
    \{\delta g_{\text{\tiny{A}} \mu}\ell^\mu\}
  +
  \hat{\nabla}_{\text{\tiny{A}}} \mathring{\nabla}_{\ell}
    \{\delta g_{\text{\tiny{B}}\mu}\ell^\mu\}
  -
  r^{-2} \mathring{\nabla}_{\ell}\{r^{2}\mathring{\nabla}_{\ell}\;\delta  g_{\text{\tiny{AB}}}\}
  -
  \hat{\nabla}_{\text{\tiny{B}}}\hat{\nabla}_{\text{\tiny{A}}}
  \{ \delta g_{\mu \nu}\ell^\mu\ell^\nu\}
  +
  \frac{1}{r}\left(\hat{\nabla}_{\text{\tiny{B}}} \left
      \{\delta g_{\text{\tiny{A}} \mu}\ell^\mu\right\} 
    +
    \hat{\nabla}_{\text{\tiny{A}}}\left\{
      \delta g_{\text{\tiny{B}}\mu}\ell^\mu\right\}\right)~~~~~~ \\
  - 
  \half\left(2\hat{\nabla}^{\text{\tiny{C}}}
    \mathring{\nabla}_{\ell} \{\delta g_{\text{\tiny{C}} \mu}\ell^\mu
    \}
    -
    r^{-2} \mathring{\nabla}_{\ell}\{r^{2}\mathring{\nabla}_{\ell}\;\delta  g_{\;\text{\tiny{C}}}^{\text{\tiny{C}}}\}
    -
    \hat{\nabla}_{\text{\tiny{C}}}
    \hat{\nabla}^{\text{\tiny{C}}}\{\delta g_{\mu \nu}\ell^\mu
    \ell^\nu\}
    +
    \frac{2}{r}\hat{\nabla}^{\text{\tiny{C}}}\{\delta
    g_{\text{\tiny{C}} \mu}\ell^\mu\}\right)r^2 \hat{g}_{\text{\tiny{AB}}} 
  \bigr].~~~~~~~~~~~~~~~~~~~~~~~~~~~~~~~~~~~~~~
\end{split}
\end{equation}
Specialization to odd and even parity GW metric
perturbations, given the auxiliary variables $\text{w}_{k\ell m}^{(\pm)}$
and the vacuum linearized Einstein equations, is shown in
sections~\ref{weyl-odd} and~\ref{weyl-even}. The expressions then
obtained for odd and even parity $\omega_{\text{\tiny{AB}}}^{\pm}$are
used as the target values
$\omega^{\pm}_{\text{\tiny{AB}}}|_{\text{\tiny{BC}}}$.

\paragraph{Odd Parity}
\label{weyl-odd}

Substituting the odd parity metric perturbations of Eq.~(\ref{e:OddDecomp}) into
Eq.~(\ref{e:first-line-simp}), one obtains
\begin{equation}
  \label{e:odd_weyl_before_simp}
  \begin{split}
    \omega^-_{\text{\tiny{AB}}} =
    \half
    \bigl[\hat{\nabla}_{\text{\tiny{B}}}\mathring{\nabla}_{\ell}\{
      h_a\ell^aS_{\text{\tiny{A}}}\}
    +
    \hat{\nabla}_{\text{\tiny{A}}}\mathring{\nabla}_{\ell}\{h_a\ell^aS_{\text{\tiny{B}}}\}
    -
    r^{-2} \mathring{\nabla}_{\ell}
    \bigl(r^{2}\mathring{\nabla}_{\ell}  \{2kS_{\text{\tiny{AB}}}\}\bigr)
     +
    \frac{1}{r}\bigl(\hat{\nabla}_{\text{\tiny{B}}}\{h_a\ell^aS_{\text{\tiny{A}}}\}
    +
    \hat{\nabla}_{\text{\tiny{A}}}\{h_a\ell^aS_{\text{\tiny{B}}}\}\bigr)~~~~~~~~~~~~~~~~~~~~\\
    -
    \half\left(2 \hat{\nabla}^{\text{\tiny{C}}}
      \mathring{\nabla}_{\ell} \{h_{a}\ell^aS _{\text{\tiny{C}}}\} 
      -
      r^{-2} \mathring{\nabla}_{\ell}\bigl(r^{2}\mathring{\nabla}_{\ell}\{2kS_{\;\text{\tiny{C}}}^{\text{\tiny{C}}}\}\bigr)
    +
    \frac{2}{r}\hat{\nabla}^{\text{\tiny{C}}}\{h_{a}\ell^aS _{\text{\tiny{C}}}\}\right)r^2
    \hat{g}_{\text{\tiny{AB}}}\bigr].~~~~~~~~~~~~~~~~~~~~~~~~~~~~~~~~~~~~~~~~~~~~~~~~~~~~~~~~~~~
  \end{split}
\end{equation}
Expansion of the last (trace) term of
Eq.~(\ref{e:odd_weyl_before_simp}) yields terms involving
$\hat{\nabla}_{\text{\tiny{C}}}(h_a \ell^a)$,
$\hat{\nabla}_{\text{\tiny{C}}} S^{\text{\tiny{C}}}$ and
$S_{\text{\tiny{C}}}^{\;\text{\tiny{C}}}$ which equal zero with the
result that this last term vanishes.  Further expansion and subsequent
simplification of Eq.~(\ref{e:odd_weyl_before_simp}) gives
\begin{equation}
  \label{e:FirstSimplification}
  \omega^-_{\text{\tiny{AB}}} =\left[D\left(h_a \ell^a\right)
    +\frac{2}{r} D k-\frac{2}{r^2} k
    -D^2k\right]S_{\text{\tiny{AB}}}.
\end{equation}
The auxiliary variables as defined in Eq.~(\ref{e:AuxVars}), are now
introduced through the term $D^2k$. Using Eq.~(45) of~\cite{Rinne2009}
to relate $Dk$ to the auxiliary variables and using Eq.~(8)
of~\cite{Rinne2009} to find expressions for $D \text{w}_{0 \ell m}^-$
and $D \text{w}_{1 \ell m}^-$, we obtain
\begin{equation}
  \label{e:SecondSimplification}
  D^2k =D\left(h_a \ell^a\right)
    +\frac{2}{r} D k -\frac{2}{r^2}k+r^2 \text{w}_{2 \ell m}^-.
\end{equation}
One can see by inspection that substituting Eq.~(\ref{e:SecondSimplification}) into
Eq.~(\ref{e:FirstSimplification}) gives the final simple and elegant
result for odd parity:
\begin{equation}
  \label{e:WeylOdd}
  \omega^-_{\text{\tiny{AB}}}=-r^2\text{w}_{2 \ell m}^-\;S_{\text{\tiny{AB}}}.
\end{equation}
This value will now be used for the odd parity target value,
$\omega^-_{\text{\tiny{AB}}}|_{\text{\tiny{BC}}}$.

\paragraph{Even Parity}
\label{weyl-even}

Specializing Eq.~(\ref{e:first-line-simp}) to even parity using the
perturbations given in Eq.~(\ref{e:EvenDecomp}), one obtains
\begin{eqnarray}
  \label{even_weyl_before_simp}
  \omega_{\text{\tiny{AB}}} &=& \half \; \bigl[
  \hat{\nabla}_{\text{\tiny{B}}}\left(\mathring{\nabla}_{\ell}
  \{ \ell^a Q_{a} Y_{\text{\tiny{A}}}\}\right)
  +
  \hat{\nabla}_{\text{\tiny{A}}} \left(\mathring{\nabla}_{\ell}\{ \ell^a Q_{a} Y_{\text{\tiny{B}}}\}\right)  
  -
   r^{-2} \mathring{\nabla}_{\ell}\bigl(r^{2}\mathring{\nabla}_{\ell}\{K
  \hat{g}_{\text{\tiny{AB}}}\;Y+G\;Y_{\text{\tiny{AB}}}\}\bigr)
  \\ \nonumber
 &&-
  \hat{\nabla}_{\text{\tiny{B}}}\hat{\nabla}_{\text{\tiny{A}}}
  \left\{H_{ab}\ell^a\ell^b\;Y\right\}
  +\frac{1}{r}\left(\hat{\nabla}_{\text{\tiny{B}}}
  \left\{\ell^a Q_{a} Y_{\text{\tiny{A}}}\right\}
  +
  \hat{\nabla}_{\text{\tiny{A}}}\left\{\ell^a Q_{a} Y_{\text{\tiny{B}}}\right\}\right)
  \bigr]^{\tt TT}.
\end{eqnarray}
Note here that it was found to be more straightforward to take the
transverse-traceless part of each term rather than to explicitly
subtract the trace as in Eq.~(\ref{e:first-line-simp}). Doing so and
simplifying gives
\begin{equation}
\label{even_weyl_simp}
  \omega_{\text{\tiny{AB}}} = 
  \half\left[2D(\ell^a Q_{a}) -D(r^2DG)
    -H_{ab}\ell^a\ell^b \right]Y_{\text{\tiny{AB}}}
\end{equation}
To make things simpler, we turn to the generalized Regge-Wheeler gauge
with the idea that if the results are true in one gauge, they are true
in all gauges since the expression is gauge invariant.  The
generalized Regge-Wheeler gauge is defined so that $Q_b$ and $G$ of
Eq.~(\ref{e:EvenDecomp}) vanish (see~\cite{Sarbach2001} page 4). Also,
in this gauge, $H_{ab}$ and $K$ correspond with their gauge-invariant
counterparts, $H^{(inv)}_{ab}$ and $K^{(inv)}$ for $\ell>2$. Thus in
the Regge-Wheeler gauge and for $\ell>2$, Eq.~(\ref{even_weyl_simp})
simplifies to
\begin{equation}
  \label{omega_even}
  \omega_{\text{\tiny{AB}}} =-\half\;H^{(inv)}_{ab}\ell^a
  \ell^b\;Y _{\text{\tiny{AB}}}.
\end{equation}
In order to introduce the auxiliary variables $\text{w}_{2 \ell m}^+$,
we make use of Eq. (19) of~\cite{Buchman_2007}, which relates
$H^{(inv)}_{ab}$ to the even parity Zerilli scalar $\Phi_{\ell m}^+$
via
\begin{equation}
  H^{(inv)}_{ab}=2\bigl(\tilde{\nabla}_a \tilde{\nabla}_b-
  \half\tilde{g}_{ab}\tilde{\nabla}^d \tilde{\nabla}_d\bigr)r\Phi_{\ell m}^+.
\end{equation}
Again, the sign difference from Eq. (19) of~\cite{Buchman_2007} makes
the sign of $\Phi_{\ell m}^+$ agree with the usual sign conventions in
numerical relativity, as mentioned earlier. Plugging into
Eq.~(\ref{omega_even}), one gets
\begin{equation}
  \label{omega_RWZscalar}
  \omega_{\text{\tiny{AB}}} =-\left[ \bigl(\tilde{\nabla}_a \tilde{\nabla}_b-
    \half\tilde{g}_{ab}\tilde{\nabla}^d \tilde{\nabla}_d\bigr)r\Phi_{\ell m}^+\right]\ell^a
  \ell^b\;Y _{\text{\tiny{AB}}},
\end{equation}
which simplifies to
$\omega_{\text{\tiny{AB}}} =-D^2(r\Phi_{\ell m}^+) Y
_{\text{\tiny{AB}}}$ noting that $\tilde{g}_{ab}\ell^a \ell^b=0$ for
Minkowski spacetime. This expression is gauge invariant provided the
derivatives of $\Phi_{\ell m}^+$ are gauge invariant. Using the
equation for $\text{w}^+_{2 \ell m}$ obtained from
Eq.~(\ref{e:AuxVars}), namely
\begin{equation}
r^2\text{w}^+_{2 \ell m}=r^{-1}D\bigl[r^2D \Phi_{\ell m}^+\bigr]=D^2(r\Phi_{\ell m}^+),
\end{equation}
one again arrives at a wonderfully simple result:
\begin{equation}
  \omega^+_{\text{\tiny{AB}}} =-r^2\text{w}^+_{2 \ell m}\; Y _{\text{\tiny{AB}}}.
\end{equation}
This is the value that will be used for the even parity target value,
$\omega^+_{\text{\tiny{AB}}}|_{\text{\tiny{BC}}}$.

The complete $\omega_{\text{\tiny{AB}}}$ will be the sum of odd and
even parity components for each mode.  Combining all of the above
results, we can write
\begin{equation}
  \omega_{\text{\tiny{AB}}}
  = \sum_{\ell m} -r^2(\text{w}^+_{2 \ell m}\; Y _{\text{\tiny{AB}}}
  + \text{w}_{2 \ell m}^-\;S_{\text{\tiny{AB}}})
\end{equation}

\subsubsection{dtHOBC Formulation}
\label{TimeDeriv}

In this section, the HOBCs are implemented via Eq.~(\ref{e:Eq-69}),
where $\dot{h}_{\alpha \beta}=\partial_t u^{1-}_{\alpha \beta}$. The
result is
\begin{equation}
  d_t u^{1-}_{\text{\tiny{AB}}} 
 \; \hat{=}\;
            P^\mathrm{P}_{\text{\tiny{AB}}}{}^{\gamma\delta}\left(\partial_t{u}^{1-}_{\gamma\delta}\right)
            = -P^\mathrm{P}_{\text{\tiny{AB}}}{}^{\gamma\delta} r^2 (\partial_t^2 + \partial_t\partial_r)
            (r^{-2} \delta g_{\gamma\delta})-\gamma_2 
            P^\mathrm{P}_{\text{\tiny{AB}}}{}^{\gamma\delta}\;\partial_t \delta g_{\gamma\delta}.
 \label{e:time-derivativeBc}
\end{equation}
Notice that the $\gamma_2$ term is annihilated by
$P^\mathrm{P}_{\alpha\beta}{}^{\gamma\delta}$ in the algebraic
boundary condition implemented in~\cite{Rinne2009}, but not in the
corresponding time-derivative boundary condition.

Since $P_{\text{\tiny{AB}}}=r^2 \hat{g}_{\text{\tiny{AB}}}$, it
follows that
$P^\mathrm{P}_{\text{\tiny{AB}}}{}^{\gamma\delta} =
P_{\text{\tiny{A}}}{}^\gamma P_{\text{\tiny{B}}}{}^\delta -( r^2/2)
\hat{g}_{\text{\tiny{AB}}} P^{\gamma\delta}$. Also keeping in mind
that $P^{\text{\tiny{AB}}}=r^{-2} \hat{g}^{\text{\tiny{AB}}}$ and
$P_{\text{\tiny{A}}}^{\, {\text{\tiny{B}}}}= \delta_{\text{\tiny{A}}}^{\, {\text{\tiny{B}}}}$, we obtain
\begin{eqnarray}
  d_t u^{1-}_{\text{\tiny{AB}}}&\hateq& -r^2 (\partial_t^2 + \partial_t\partial_r)
                                        (r^{-2} \delta
                                        g_{\text{\tiny{AB}}}^{\tt
                                        TT})-\gamma_2
                                        \;\partial_t \delta
                                        g_{\text{\tiny{AB}}}^{\tt
                                        TT}
\label{e:time-derivBc2-Spherical}
\end{eqnarray}
Expressions for the right-hand-side of
Eq.~(\ref{e:time-derivBc2-Spherical}) are derived in the
Regge-Wheeler-Zerilli formalism which involve the auxiliary variables
at the boundary.

\paragraph{Odd Parity}

The gauge-invariant potential $h^{(\mathrm{inv})}$~\cite{Sarbach2001}
is related to the odd parity Regge-Wheeler scalar $\Phi_{\ell m}^{-}$ via
$h^{(\mathrm{inv})} = \tilde * {\rm d}(r \Phi_{\ell m}^{-})$, where
$\tilde *$ denotes the Hodge dual with respect to $\tilde g$.  In
particular, $\tilde * u_a dx^a=\tilde{\epsilon}_{ab}u^adx^b$ (from the
top of page 3 of \cite{Sarbach2001}), where $\tilde{\epsilon}_{ab}$ is
the standard volume element in $(\tilde{M},\tilde{g})$, oriented so
that $\tilde{\epsilon}_{tr}=|\tilde{g}|^{1/2}$. For flat spacetime,
$\tilde{\epsilon}_{tr}=-\tilde{\epsilon}_{rt}=1$, and
$\tilde{\epsilon}^{\,tr}=-\tilde{\epsilon}^{\,rt}=-1$. Applying this
to the expression
$h^{(\mathrm{inv})} = \tilde * {\rm d}(r \Phi_{\ell m}^{-})$, we obtain
\begin{equation}
  h^{(\mathrm{inv})} = ~~\tilde{\epsilon}_{ab} \;\partial^a(r \Phi_{\ell m}^{-}) dx^b
                     = -\partial_t(r \Phi_{\ell m}^{-}) dr - \partial_r(r \Phi_{\ell m}^{-}) dt.
\end{equation}
Since the 1-form $h^{(\mathrm{inv})}$ is decomposed as
$h^{(\mathrm{inv})}=h^{(\mathrm{inv})}_r dr+h^{(\mathrm{inv})}_t dt$,
it follows that
\begin{equation}
\label{e:InvMetricPert1}
   h_r^{(\mathrm{inv})} = -\partial_t (r \Phi_{\ell m}^{-})~~~~{\text{and}}~~~~
  h_t^{(\mathrm{inv})} = -\partial_r (r \Phi_{\ell m}^{-}).
\end{equation}
Using Eqs.~(16) and (17) of~\cite{Rinne2009}, which are
$h^{(\mathrm{inv})}_t = h_t - \partial_t k$ and
$h^{(\mathrm{inv})}_r = h_r - r^2 \partial_r(r^{-2}k)$, respectively,
we re-express the time-derivative boundary
condition~(\ref{e:time-derivBc2-Spherical}) in terms of the amplitudes
$h_a$ and the odd-parity auxiliary variables $\text{w}_{k \ell m}^-$
to give
\begin{equation}
(\partial_t+\partial_r)(r^{-2}k)=r^{-2}(h_t+h_r+r^2\text{w}_{1 \ell
  m}^-+r\text{w}_{0 \ell m}^-).
\end{equation}
With
$\delta g_{\text{\tiny{AB}}}^{\tt
  TT}=2kS_{\text{\tiny{AB}}}$~\cite{Rinne2009}, substitution into
Eq.~(\ref{e:time-derivBc2-Spherical}) gives
\begin{equation}
\label{e:time-derivBc3-Spherical-odd}
d_t u^{1-}_{\text{\tiny{AB}}}\hateq
-2[\dot{h}_t+\dot{h}_r+r^2\dot{\text{w}}_{1 \ell
  m}^-+r\dot{\text{w}}_{0 \ell m}^-+\gamma_2 \;\dot{k}]S_{\text{\tiny{AB}}}.
\end{equation}
The time derivative $\dot{\text{w}}_{0 \ell m}^-$ is removed by
recalling $r\text{w}_{0 \ell m}^-=\Phi_{\ell m}^{-}$ and
$\dot{\text{w}}_{1 \ell m}^-$ using Eq.~(\ref{e:AuxODEs}).  The final
result for odd parity in the dtHOBC formulation is:
\begin{equation}
\label{e:time-derivBc4-Spherical-odd}
d_t u^{1-}_{\text{\tiny{AB}}}\hateq
-2\left[\dot{h}_t+\dot{h}_r-\half\ell(\ell+1)\text{w}_{0 \ell
    m}^--r\text{w}_{1 \ell m}^-+\half r^2\text{w}_{2 \ell m}^-
  -\frac{1}{r}\left(h_r+2k/r-\partial_r k\right)
  +\gamma_2 \;\dot{k}\right]S_{\text{\tiny{AB}}}
\end{equation}
for $\ell \ge2$. All the time derivatives of the metric perturbations
($\dot{h}_t$, $\dot{h}_r$, and $\dot{k}$) are calculable in {\tt
  SpEC}.

\paragraph{Even Parity}

Following~\cite{Sarbach2001} (page 5), a scalar field $\zeta$ is
introduced according to $Z = d \zeta$, where $Z$ is the Zerilli
one-form. The Zerilli scalar is then defined as
${\Phi_{\ell m}^{+}} \equiv -\zeta/\Lambda$, recalling that
$\Lambda \equiv (\ell-1)(\ell+2)$.  Since $Z=Z_r dr + Z_t dt$ and
$d\zeta = \partial_r \zeta dr + \partial_t \zeta dt$, it follows that
$Z_r=\partial_r \zeta$ and $Z_t=\partial_t \zeta$. Thus,
\begin{equation}
\label{e:Z1}
Z_t + Z_r = -\Lambda (\partial_t + \partial_r)  \Phi_{\ell m}^{+} =
-\Lambda r \text{w}_{1 \ell m}^+.
\end{equation}
The Zerilli one-form of Eq.~(\ref{e:Z1}) is combined with the
alternate form obtained using the amplitudes of the even parity metric
perturbations to give
\begin{equation}
  \label{e:EvenBdryData1}
  r^2 (\partial_t + \partial_r) G = 2 r^2 \text{w}_{1 \ell m}^{+} 
  - \tfrac{2}{\Lambda} r^2 (\partial_t + \partial_r) K +
  \tfrac{2}{\Lambda} r (\dot Q_r - \partial_r Q_t - \tfrac{2}{r} Q_r + H_{tr} + H_{rr}).
\end{equation}
(Note that this equation was also derived in Sec.~2.5.2
of~\cite{Rinne2009} modulo the sign in front of
$\text{w}_{1 \ell m}^{+}$). The substitution of
$\delta g_{\text{\tiny{AB}}}^{\tt TT} = r^2 G Y_{\text{\tiny{AB}}}$
into Eq.~(\ref{e:time-derivBc2-Spherical}) gives
\begin{eqnarray}
  d_t u^{1-}_{\text{\tiny{AB}}}
  &\hateq& - \partial_t [r^2 (\partial_t + \partial_r)\;G] Y_{\text{\tiny{AB}}}
           -\gamma_2  r^2 \;\dot{G} Y_{\text{\tiny{AB}}}.
\end{eqnarray}
Substituting for the expression for $r^2 (\partial_t + \partial_r) G$
given in Eq.~(\ref{e:EvenBdryData1}) and eliminating
$\dot{\text{w}}_ {1 \ell m}^+$ using Eq.~(\ref{e:AuxODEs}), we obtain
the final result for even parity, namely:
\begin{equation}
  \label{e:time-derivBc4-Spherical-even}
  d_t u^{1-}_{\text{\tiny{AB}}}\hateq  [\ell(\ell+1)\text{w}_{0 \ell
    m}^++2r\text{w}_{1 \ell m}^+-r^2\text{w}_{2 \ell m}^+
  +\tfrac{2}{\Lambda} r^2(\ddot{K}+\partial_r \dot{K})
  -\tfrac{2}{\Lambda} r (\ddot Q_r - \partial_r \dot{Q}_t 
  - \tfrac{2}{r} \dot{Q}_r + \dot{H}_{tr} + \dot{H}_{rr}) -\gamma_2  r^2 \;\dot{G}] Y_{\text{\tiny{AB}}}
\end{equation}
for $\ell \ge 2$.

\section{Results}
\label{results}

\subsection{Multipolar Waves}
\label{MPwaves}

We first evolve GW packets on a flat background to
test our new boundary conditions.  The tests we perform are
similar to those in~\cite{Rinne2009}. We evolve a solution of the
linearized (about flat spacetime) Einstein equations, but using the
fully nonlinear {\tt SpEC} code.  We will compare the numerical
solution to the (linearized) analytic solution, and we expect
differences of order $A^2$, where $A$ is the amplitude of the
linearized solution. For the evolution, we choose the harmonic gauge,
which is consistent with the analytic solution.

The analytic solution is given by Rinne~\cite{Rinne2009b}, and is a
generalization of the ($\ell=2$) Teukolsky wave~\cite{Teukolsky_1982}
to arbitrary multipoles. The solution depends on freely specifiable
mode functions $F(r-t)$, where $(r-t)$ is a retarded radius. We choose
$F$ with the same form as in~\cite{Rinne2009}:
\begin{equation}
  \label{e:GaussianProfileFct}
  F(r-t) = A \exp \left[ -\frac{(r-r_0-t)^2}{\sigma^2} \right].
\end{equation}
Here, $A$ is the amplitude, $\sigma$ is the width, and $r_0$ is the
radius of the peak of the wave packet at $t=0$.  The flat spacetime
master equation for the Regge-Wheeler-Zerilli scalars,
Eq.~(\ref{e:FlatRWZequation}), can be solved by making the ansatz
\begin{equation}
\Phi_{\ell m}^{(\pm)}=\sum_{j=0}^\ell c_{j\;(\ell,m)}r^{j-\ell}F^{(j)}(r-t),
\end{equation}
where $F^{(j)}(r-t)\equiv d^j F(r-t)/d(r-t)^j $for an outgoing
solution.  For all reported multipolar wave tests, the outer boundary
radius $R_{\rm bdry}$ of the computational domain is placed at
$R_{\rm bdry}=30$, and the width of the Gaussian wave-packet is chosen
to be $\sigma=1.5$.

In addition to the wave-packet envelope function $F$, a multipolar
wave is characterized by its angular mode numbers ($\ell,m$) and and
by its parity, even (+) or odd (-).  Furthermore, each mode has
distinct real and imaginary components (constrained by the requirement
that $\Phi_{\ell m}^{(\pm)}$ itself be real). We have tested our
boundary condition algorithms with both $\ell=3$ and $\ell=4$ modes
and both even and odd parity, with very similar results in all cases.
In addition, we have evolved a multipolar wave whose center is
displaced from the coordinate origin, so that wavefronts do not match
coordinate spheres on the outer boundary, and the wave is a mixture of
modes in the tensor spherical harmonic decomposition in our code's
coordinates.  Results for such displaced waves are also very similar
to those for centered waves.  Therefore, we choose to pick one
representative case and provide an exhaustive description of its
behavior under different initialization and evolution choices.  We
choose an $\ell=4$, $m=2$, even parity wave.  Given the usual complex
spherical harmonics, the reality condition for $\Phi^{(\pm)}_{\ell m}$
requires
$\left(\Phi^{(\pm)}_{\ell, m}\right)^{\ast} =
(-1)^m\Phi^{(\pm)}_{\ell, -m}$, so in this basis an appropriate
$\ell=4$, $m=-2$ mode is also present.

A final choice is whether to evolve in an asymptotically Minkowski or
in a rotating coordinate system ({\it ie.} a non-rotating or a rotating
grid).  For this particular problem, there is no advantage to the
latter (and no natural non-zero choice for angular velocity). However,
since many of the most interesting applications involve binaries, for
which {\tt SpEC} uses co-rotating frames, we have tried evolving with
angular speed $\Omega=0.01$ (so equatorial boundary grid points move
at 0.3$c$).  This has negligible effect on the results, except that
the problematic late-time behavior for the dtHOBC implementation
discussed below is only seen in runs with rotating grid.  Therefore,
we report only results for runs with rotating grids below.

\subsubsection{Evolution of an interior wave}
\label{EvolnInterWave}

We first evolve a wave with $r_0=15$, so the wave begins in the
interior far from the boundary and propagates outward.  For this
choice, the wave will vanish at the boundary to numerical roundoff
error at $t=0$, so it is acceptable to initialize
$\text{w}_{k\ell m}^{(\pm)}=0$.  We then track the RWZ scalar
$\Phi^{(\pm)}_{\ell m}$ in the subsequent evolution and compare it to
$\Phi^{(\pm)}_{\ell m}$ for the analytic solution of the linearized
equations.  The difference between the two, denoted
$\Delta \Phi^{(\pm)}_{\ell m}$, is initially zero (by construction)
but will become nonzero at later times for three reasons.

First, there can be reflections from the outer boundary due to
imperfect boundary conditions.  This is the deviation that interests
us, which the HOBCs are designed to minimize.  The second is numerical
truncation error; however, convergence testing shows that this is not
a major contribution to the difference between numerical and analytic
solutions.  The third reason is that the numerical and analytic
solutions are solutions to a different set of equations --- to the
full and linearized Einstein equations, respectively.  This is a large
effect, but it can be controlled by varying the amplitude, $A$, of the
wave packet.  Differences due to reflections can scale as $A$, while
differences due to nonlinearities in Einstein's equations must scale
as $A^2$ or higher.  Therefore, following earlier
work~\cite{Rinne2009}, we perform convergence tests in $A$, plotting
$A^{-1}\Delta \Phi^{(\pm)}_{\ell m}$.  In the absence of reflections,
$A^{-1}\Delta \Phi^{(\pm)}_{\ell m}$ will decrease as $A$ decreases
until a roundoff error floor is reached.

\begin{figure}[!htb]
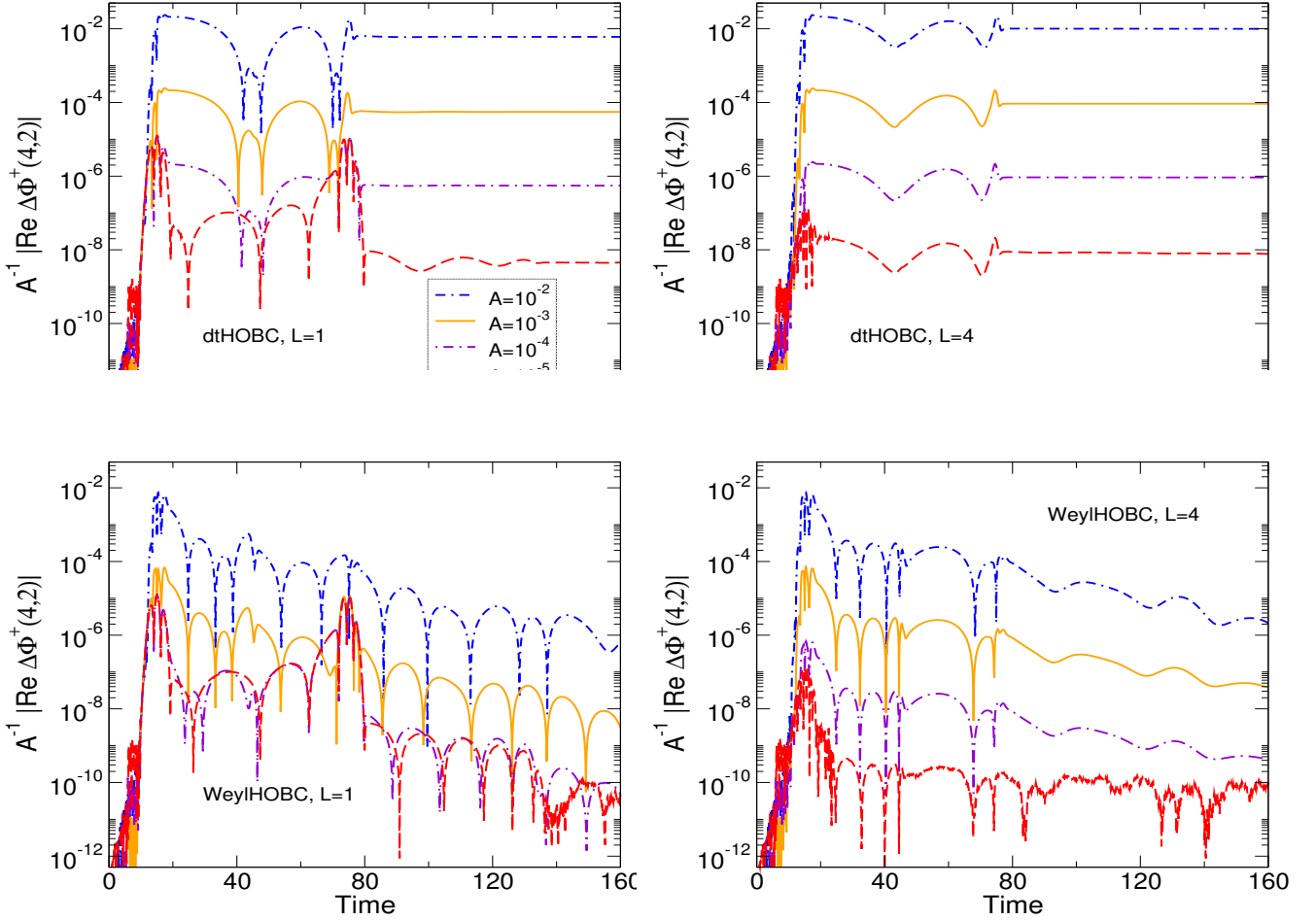

\includegraphics[width=8.75cm, height=6.25cm]{Figure1a.eps}
\includegraphics[width=8.75cm, height=6.25cm]{Figure1b.eps}
\includegraphics[width=8.75cm, height=6.25cm]{Figure1c.eps}
\includegraphics[width=8.75cm, height=6.25cm]{Figure1d.eps}
\caption{Amplitude convergence of even parity multipolar wave with
  (4,2) mode. The wave is initially centered in the interior with
  $r_0=15$. The $\text{w}_{k\ell m}^{(\pm)}$'s are initialized to
  zero. Upper left: dtHOBC formulation with $L=1$. Upper right: dtHOBC
  with $L=4$. Lower left: WeylHOBC formulation with $L=1$. Lower
  right: WeylHOBC with $L=4$.}
\label{f:r15-wk-zero}
\end{figure}

This is, in fact, what happens for these interior wave evolutions, for
both dtHOBC and WeylHOBC implementations, as shown in
Figure~\ref{f:r15-wk-zero}.  We show results for boundary condition
orders $L=1$ and $L=4$.  For an $\ell=4$ wave, there should be
reflections for boundary condition order $L=1$
(see~\cite{Buchman_2006}) but not for order $L=4$.  These reflections
should manifest as contributions to $\Delta \Phi^{(\pm)}_{\ell m}$
linear in $A$, so that $A^{-1}\Delta \Phi^{(\pm)}_{\ell m}$ should
converge to zero for the 4th order boundary condition runs, but it
should converge to a nonzero function for the 1st order boundary
condition runs.  This can be seen especially by looking at the early
evolution $t<80$; the separation of resolutions (indicating
convergence to zero) is clearly much cleaner for the output of the
$L=4$ runs. For the $L=1$ runs, one sees that
$A^{-1}\Delta \Phi^{(+)}_{4,2}$ converges to two pulses at the early
evolution as $A$ decreases. The first pulse, at a time of $t=15$,
occurs when the outgoing wave reaches the boundary (and is also seen
in the $L=4$ runs). The second pulse, at $t=75$, is the result of a
spurious reflection occurring at $t=15$ in the $L=1$ run. The time
interval between the two pulses, $\Delta t = 60$, is the time needed
for this reflection to travel inward across the Cauchy domain (which
is a radial distance of $30$), and outward again to the outer
boundary. Recall that, for the WeylHOBC implementation, the 1st-order
boundary condition is algorithmically identical to the
freezing-$\Psi_0$ boundary condition usually used by {\tt SpEC}, but
for the dtHOBC implementation, the 1st-order boundary condition is
only identical to {\tt SpEC}'s standard boundary condition to
truncation error.

\begin{figure}[!htb]
\includegraphics[width=8.75cm, height=6.25cm]{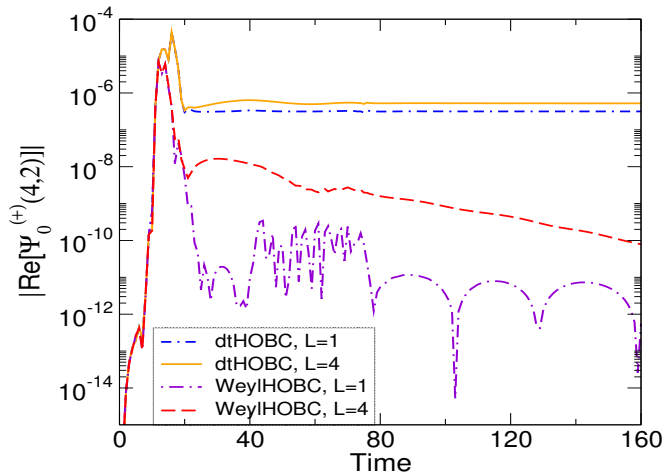}
\caption{Real $\Psi^+_0$ (4,2) mode. The wave is initially centered in
  the interior with $r_0=15$. The wave amplitude is $10^{-2}$ in all
  cases.}
\label{f:r15-Psi0} 
\end{figure}

A notable feature at late times is that, for each run,
$A^{-1}\Delta \Phi^{(\pm)}_{\ell m}$ settles to a nonzero constant.
This is a nonlinear effect, as demonstrated by the fact that the
settled value converges to zero as amplitude is decreased.
Nevertheless, it indicates that, even after the wave is long past, the
interior spacetime plus boundary system may settle to some stationary
vacuum state other than Minkowski.  This possibly could be a gauge
effect so that the system settles to flat spacetime in some slightly
different coordinate system.  We test this by plotting the
Newman-Penrose scalar $\Psi_0$.  As a projection of the Weyl tensor,
$\Psi_0$ should go to zero if the spacetime is settling to a
zero-curvature state.  Figure~\ref{f:r15-Psi0} shows $\Psi_0^+(4,2)$
for the high-amplitude case $A=10^{-2}$.  We see that $\Psi_0^+(4,2)$
does decrease toward zero at late times for WeylHOBC but not for
dtHOBC.  Perhaps this is because the dtHOBC implementation only
controls the time derivative of the metric on the boundary and not the
metric itself.

\subsubsection{Initial wave on the boundary:  early times}
\label{r30-early}

\begin{figure}[!htb]
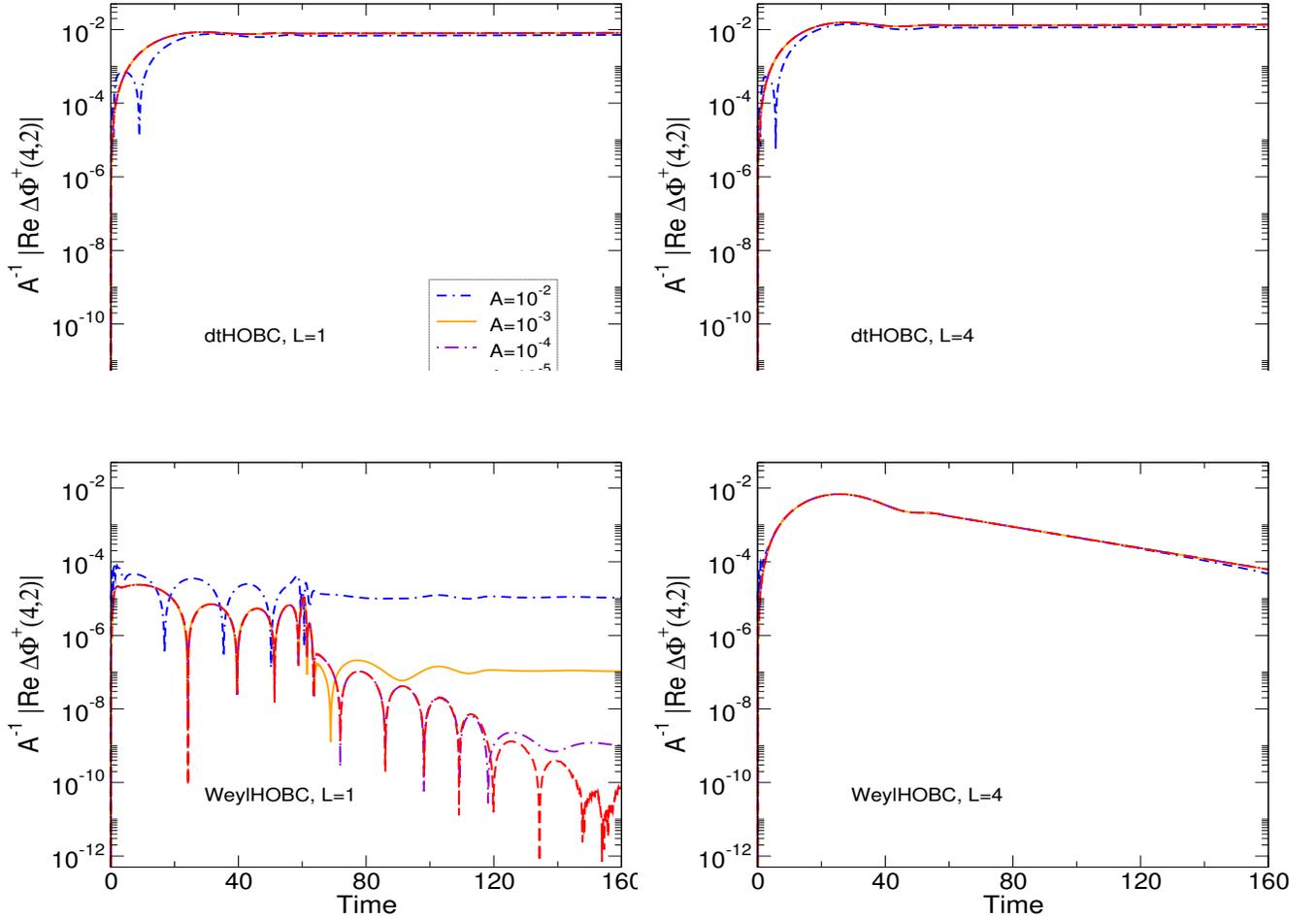

\includegraphics[width=8.75cm, height=6.25cm]{Figure3a.eps}
\includegraphics[width=8.75cm, height=6.25cm]{Figure3b.eps}
\includegraphics[width=8.75cm, height=6.25cm]{Figure3c.eps}
\includegraphics[width=8.75cm, height=6.25cm]{Figure3d.eps}
\caption{Same as Figure~\ref{f:r15-wk-zero} except $r_0=30$.}
\label{f:dtHOBC-r30-wk-zero}
\end{figure}

Next we center the wave on the boundary, with $r_0=30$. In this case,
initialization of $\text{w}_{k\ell m}^{(\pm)}$ is nontrivial. With the
wave centered initially on the boundary, initializing
$\text{w}_{k\ell m}^{(\pm)} = 0$ will produce spurious
reflections. One can see by studying
Figure~\ref{f:dtHOBC-r30-wk-zero}, that the
$A^{-1} \Delta \Phi^{(\pm)}_{\ell m}$ convergence fails; thus, we do
not recover the analytic solution at early times.  This indicates that
there are linear errors corresponding to reflections.  The exception
is the first-order Weyl implementation, which does not show strong
early-time reflections.  This is expected because order $L=1$ WeylHOBC
is identical to the standard freezing-$\Psi_0$ boundary condition for
which the auxiliary system is not used; thus, it does not matter how
poorly it is initialized.

The proper initialization of the auxiliary variables should allow
$\Phi^{(\pm)}_{\ell m}$ to evolve correctly as the wave passes through
the outer boundary. We extract $\text{w}_{k\ell m}^{(\pm)}$ values
consistent with the compatibility conditions as described
in~\ref{InitAuxVars}. In Table~\ref{tab:w_variance_table}, we compare
the extracted to the analytic solutions of the linearized Einstein
equations, and see that we can extract up to $\text{w}_ {1 \ell m}$
and $\text{w}_ {2 \ell m}$ to yield a correct answer, but not up to
$\text{w}_ {3 \ell m}$. Even apart from an analytic value with which
to compare, the variance in the fit indicates that the extracted
$\text{w}_ {3 \ell m}$ is unreliable.
\begin{table}[!htb]
  \begin{ruledtabular}
  \begin{tabular}{l|c|c|r}
  \textrm{}&
  \textrm{Analytic}&
  \textrm{Extracted}&
  \textrm{Variance}
  \\
  \colrule
  $\text{w}_{1 \ell m}$ & $1.31431\times10^{-8}$ & $1.31432\times10^{-8}$ & $3.50284\times10^{-10}$\\
  $\text{w}_{2 \ell m}$ & $-4.36397\times10^{-10}$ & $-4.36469\times10^{-10}$ & $4.95063\times10^{-12}$\\
  $\text{w}_{3 \ell m}$ & $-1.70706\times10^{-12}$ & $-1.47191\times10^{-13}$ & $8.42510\times10^{-14}$\\
  \end{tabular}
  \end{ruledtabular}
  \caption{Comparison of extracted with analytic
    $\text{w}_{k\ell m}^{(\pm)}$, including
    variance, for initialization when an even parity (4,2) multipolar
    wave is centered on the outer boundary at time $t=0$. 
    The WeylHOBC implementation is used, and the
    multipolar wave amplitude is $10^{-4}$.}
  \label{tab:w_variance_table}
\end{table}

\begin{figure}[!htb]
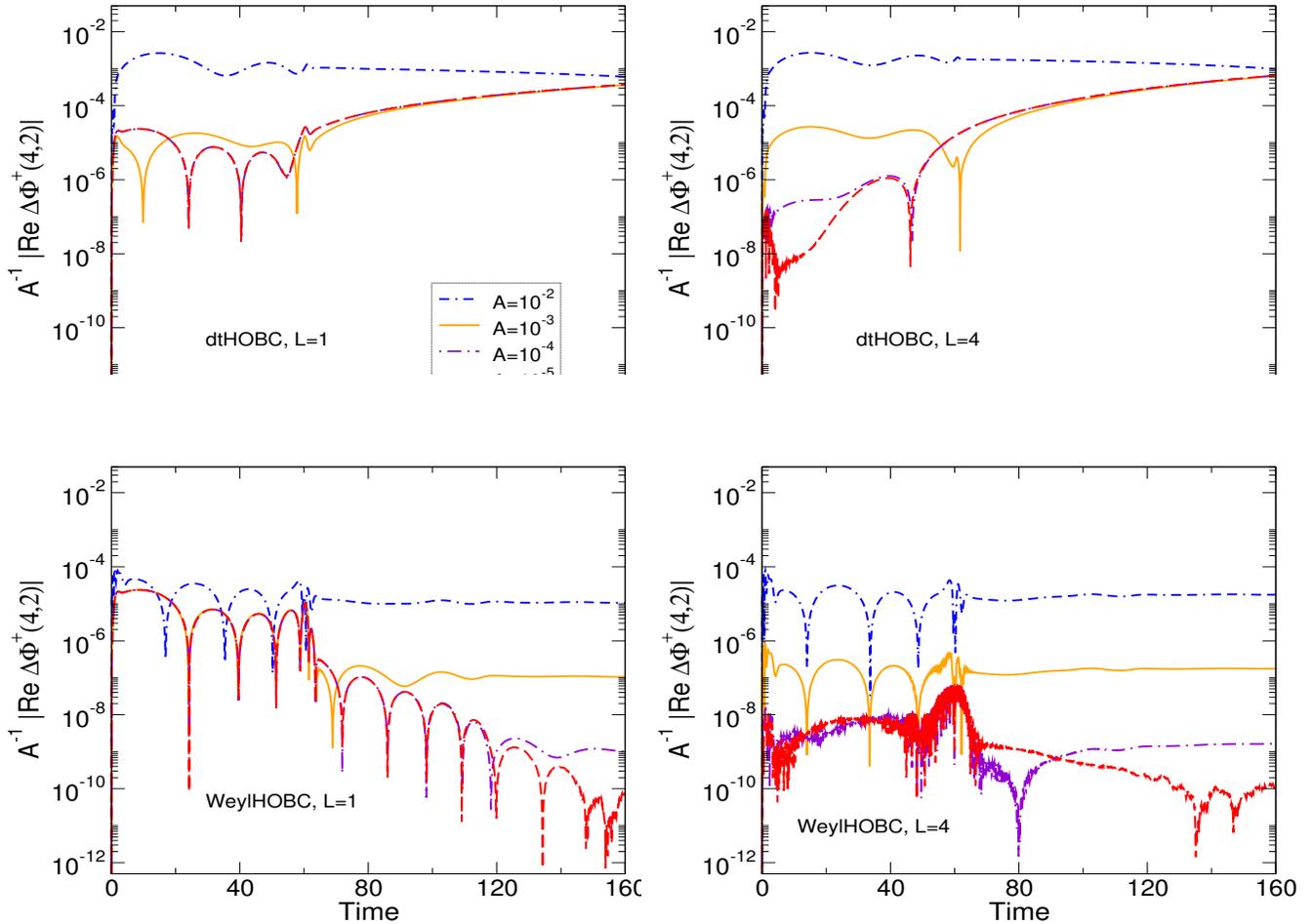

\includegraphics[width=8.75cm, height=6.25cm]{Figure4a.eps}
\includegraphics[width=8.75cm, height=6.25cm]{Figure4b.eps}
\includegraphics[width=8.75cm, height=6.25cm]{Figure4c.eps}
\includegraphics[width=8.75cm, height=6.25cm]{Figure4d.eps}
\caption{Same as Figure~\ref{f:dtHOBC-r30-wk-zero} except for
  initialization of $\text{w}_{k\ell m}^{(\pm)}$. Here, the
  $\text{w}_{k\ell m}^{(\pm)}$'s are initialized to their analytic
  values, where $k=1,2,3,4$.}
\label{f:r30-wk-analytic}
\end{figure}

\begin{figure}[!htb]
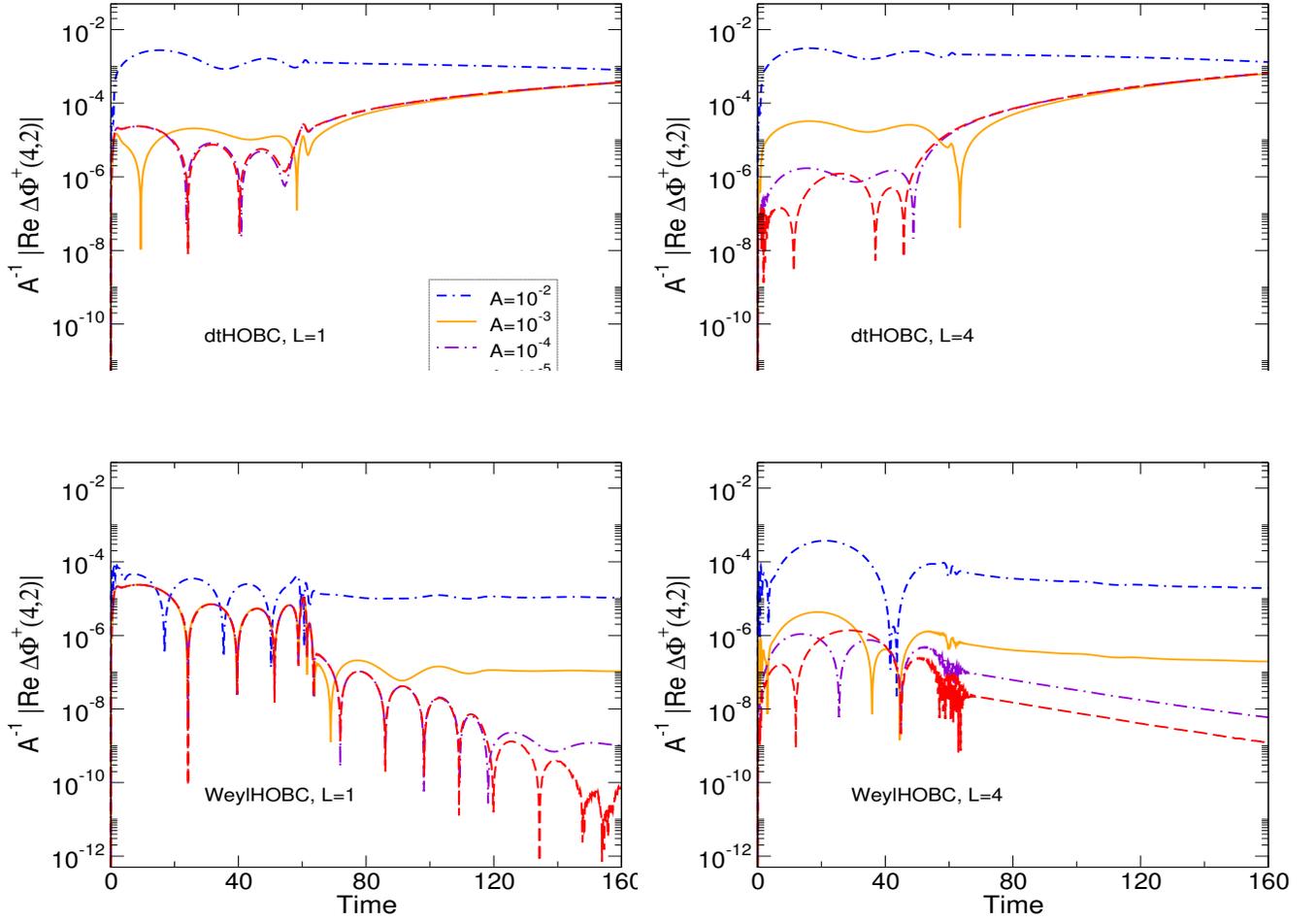

\includegraphics[width=8.75cm, height=6.25cm]{Figure5a.eps}
\includegraphics[width=8.75cm, height=6.25cm]{Figure5b.eps}
\includegraphics[width=8.75cm, height=6.25cm]{Figure5c.eps}
\includegraphics[width=8.75cm, height=6.25cm]{Figure5d.eps}
\caption{Same as Figures~\ref{f:dtHOBC-r30-wk-zero}
  and~\ref{f:r30-wk-analytic} except for initialization of
  $\text{w}_{k\ell m}^{(\pm)}$. Here the
  $\text{w}_{k\ell m}^{(\pm)}$'s are extracted, where $k=1,2$.}
\label{f:r30-wk-extracted}
\end{figure}

Results for $r_0=30$ with correct initial $\text{w}_{k\ell m}^{(\pm)}$
are shown in Figures~\ref{f:r30-wk-analytic}
and~\ref{f:r30-wk-extracted}.  In Figure~\ref{f:r30-wk-analytic}, we
initialize $\text{w}_{k\ell m}^{(\pm)}$ up through $k=4$ (the highest
used by the 4th order method) using the analytic multipolar wave
solution.  In Figure~\ref{f:r30-wk-extracted}, we initialize
$\text{w}_{k\ell m}^{(\pm)}$ using all of the reliably extracted
values, leaving the others zero.  In both cases, we see the expected
convergence with amplitude.  This is most clear for the first $t=50$
of the evolutions when reflections, if present, are visible.

\subsubsection{Initial wave on the boundary:  late times}
\label{r30-late}

It is at late times, long after the wave has passed through the outer
boundary (so there should be nothing left but static Minkowski
spacetime), that the difference between dtHOBC and WeylHOBC most
clearly manifests.  We observe a drift, nearly linear with time, in
$\Phi^{(\pm)}_{\ell m}$ for the dtHOBC evolution, but not for the
WeylHOBC evolution. This drift becomes visible from about $t=50$
onward, as seen in the upper panels of Figures~\ref{f:r30-wk-analytic}
and~\ref{f:r30-wk-extracted}.  Interestingly, while the drift does
appear in the real component of the $\ell=4$, $m=2$ mode (the only
mode initially excited), it is stronger in the $\ell=4$, $m=-2$
imaginary component (consistent with the condition that
$\Phi^{(\pm)}_{\ell m}$ be real).  The drift is insensitive to the
method of initialization of $\text{w}_{k\ell m}^{(\pm)}$ ({\it ie.},
extracted vs. analytic), and its late-time slope is linear in the wave
amplitude $A$.  This suggests an effect which initially is stimulated
by a nonlinear effect (so that modes which are zero at $t=0$ can grow)
but whose later increase is a linear effect.

By introducing the auxiliary variables, we have in effect added a new
set of constraint equations to the overall system, namely the
compatibility conditions.  At any given time,
$\text{w}_{k\ell m}^{(\pm)}$ at the boundary can be computed in two
ways: either from the outgoing null derivatives of the interior metric
evaluated on the boundary or from the evolved auxiliary system.  By
properly initializing $\text{w}_{k\ell m}^{(\pm)}$, we force these
constraints to be satisfied at $t=0$. However, at $t>0$, truncation
error is expected to cause the two solutions for
$\text{w}_{k\ell m}^{(\pm)}$ at the boundary to drift apart.  We have
tried to measure $\text{w}_{1 \ell m}^{\pm}$ from the interior metric
throughout the evolution, using the techniques described in
Section~\ref{InitAuxVars}, but with observation radii inside $r=30$.
Unfortunately, these extracted $\text{w}_{1 \ell m}^{\pm}$ values are
found to be much less accurate than those obtained using exterior
points on an enlarged grid, making confident conclusions difficult.
However, we do indeed see that the two values of
$\text{w}_{1 \ell m}^{\pm}$ seem to diverge quickly (one growing and
positive, the other growing and negative) during the drift phase of
the dtHOBC evolution.  The experience of interior (volume data)
constraints in numerical relativity is that violations tend to grow
disastrously unless the formulation is carefully chosen, with terms
proportional to the violation added back to the evolution to ``damp''
constraint violations, driving the solution towards the constraint
satisfying subspace of states.  It is possible that the auxiliary
evolution system or the boundary conditions imposed on the interior
metric could be altered in such a way as to damp violations of the
compatibility conditions in the dtHOBC formulation. However, since
WeylHOBC shows no signs of any such problem, we see no reason to
pursue this line of thought further.  As we move to binary black hole
simulations presented in the next section, we utilize exclusively the
WeylHOBC implementation.

\subsection{Binary Black Holes}
\label{BBHs}

Next, we evolve an inspiralling binary black hole system. The binary
has a high mass ratio, 7:1, ensuring a significant contribution from
subdominant modes.  The initial data is constructed from a superposed
Kerr-Schild background~\cite{Lovelace2008_KS}.  The initial binary
separation is chosen to be $27M$, where $M$ is the sum of the
Christodoulou masses of the two horizons.  Both black holes are
initially non-spinning. With this initial data scenario, the time to
merger is especially long ($t_{\rm merger}\approx 106,000M$). Hence in
this paper, we focus on the inspiral phase of the coalescence only and
plan to discuss mergers in a follow-up paper. Tests are performed with
two outer boundary locations: first, the outer boundary is placed at
$R_{\rm bdry}=250M$ and second, at $R_{\rm bdry}=500M$. These are
intentionally placed closer to the source than for typical simulations
so as to test the boundary conditions.

For binary black hole simulations, there are no analytic solutions
with which to compare.  Instead, we run reference simulations, which
have grids identical to that of the test runs inside of
$R_{\rm bdry}$, but are surrounded by extra spherical shell domains
which extend the outer boundary to either $R=2,400M$ (for test runs
with $R_{\rm bdry}=250M$) or to $R=2,646M$ (for test runs with
$R_{\rm bdry}=500M$).  The reference runs use the 4th order boundary
condition in the WeylHOBC implementation.  Any spurious reflections of
outgoing GWs from this extended outer boundary radius
should be small and should not affect the test $R_{\rm bdry}$ shell before
$t\approx 4,500M$.

We evolve the high mass ratio binary black hole initial data using the
standard freezing-$\Psi_0$ {\tt SpEC} implementation and the 4th order
WeylHOBC (corresponding to $L=1$ and $L=4$, respectively), with
boundary at $R_{\rm bdry}$.  We monitor the six RWZ scalar modes with
the highest amplitudes at the outer boundary of the test runs ({\it
  ie.} at either $R_{\rm bdry}=250M$ or $R_{\rm bdry}=500M$). These
are (in order of descending amplitude) $\Phi^+_{2,2}$, $\Phi^+_{2,0}$,
$\Phi^-_{2,1}$, $\Phi^+_{3,3}$, $\Phi^+_{3,1}$, and
$\Phi^+_{4,4}$.

For the fourth order WeylHOBC, we perform two separate evolutions: (i)
with all $\text{w}_{k\ell m}^{(\pm)}$ initialized to zero, and (ii)
with
$\text{w}_{1 2 2}^+, \text{w}_{2 2 2}^+, \text{w}_{1 2 0}^+,
\text{w}_{2 2 0}^+, \text{w}_{1 2 1}^-, \text{w}_{2 2 1}^-,
\text{w}_{1 3 3}^+, \text{w}_{2 3 3}^+, \text{w}_{1 3 1}^+,
\text{w}_{2 3 1}^+, \text{w}_{1 4 4}^+, \text{w}_{2 4 4}^+$
initialized using a short evolution on the reference grid, according
to the procedure of Section~\ref{InitAuxVars} (with the rest of the
$\text{w}_{k\ell m}^{(\pm)}$ initialized to zero). The modes
initialized in the second evolution were chosen because they were
found to have the largest early-time $\text{w}_{k\ell m}^{(\pm)}$.
For this binary black hole case, we alter the fitting formula
Eq.~(\ref{eq:Phi_A}) and only fit to a quadratic function in $\lambda$
({\it ie.} we take $\sigma_3=0$).  We find that we cannot extract
$\text{w}_{3 \ell m}^{(\pm)}$ using the usual cubic function; fitting
to a cubic function only leads to large variance on all extracted
variables, although it gives almost exactly the same
$\text{w}_{1 \ell m}^{(\pm)}$ and $\text{w}_{2 \ell m}^{(\pm)}$, and
$\text{w}_{3 \ell m}^{(\pm)}$ is not distinguished from zero.  Using
the variance of the quadratic fit, we extract
$\text{w}_{1 \ell m}^{(\pm)}$ modes (for the most significant $\ell$
,$m$) with a relative uncertainty of $10^{-5}$ and
$\text{w}_{2 \ell m}^{(\pm)}$ modes with a relative uncertainty of
$10^{-2}$--$10^{-3}$.

\begin{figure}[!htb]
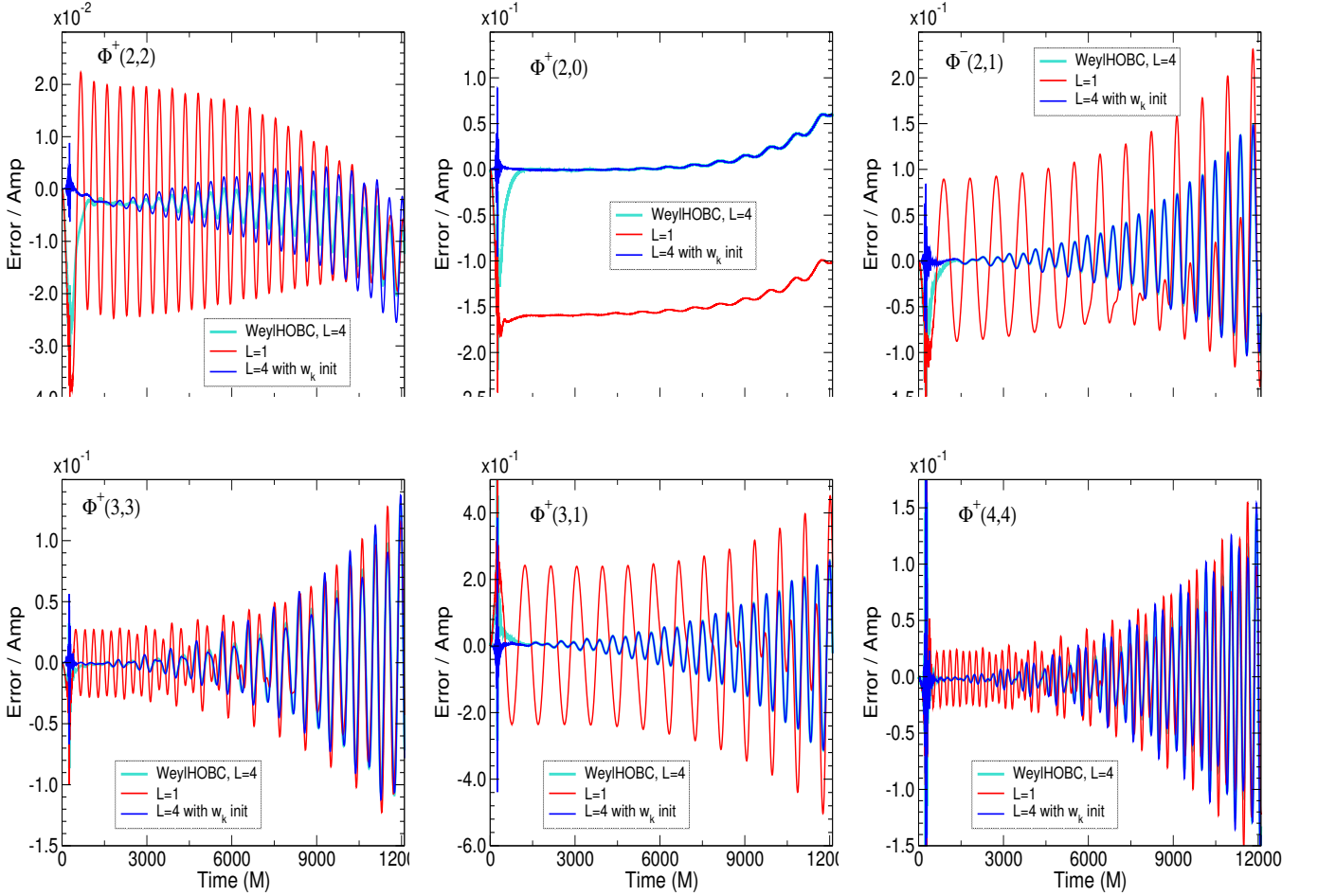

\includegraphics[width=5.9cm, height=6.25cm]{Figure6a.eps}
\includegraphics[width=5.9cm, height=6.25cm]{Figure6b.eps}
\includegraphics[width=5.9cm, height=6.25cm]{Figure6c.eps}
\includegraphics[width=5.9cm, height=6.25cm]{Figure6d.eps}
\includegraphics[width=5.9cm, height=6.25cm]{Figure6e.eps}
\includegraphics[width=5.9cm, height=6.25cm]{Figure6f.eps}
\caption{Errors for the six highest amplitude RWZ scalar modes
  (calculated and normalized as described in the text), from BBH
  simulations with test $R_{\rm bdry}=250M$ and reference
  $R_{\rm bdry}=2,400M$. All waves are extracted at $R=250M$. The
  turquoise and blue curves are runs which use the $L=4$ WeylHOBC without
  and with proper $\text{w}_{k \ell m}$ initialization,
  respectively. The red curve is a run which uses the $L=1$
  freezing-$\Psi_0$ boundary condition.}
\label{f:bbh-Rbdry250-errors}
\end{figure}

\begin{figure}[!htb]
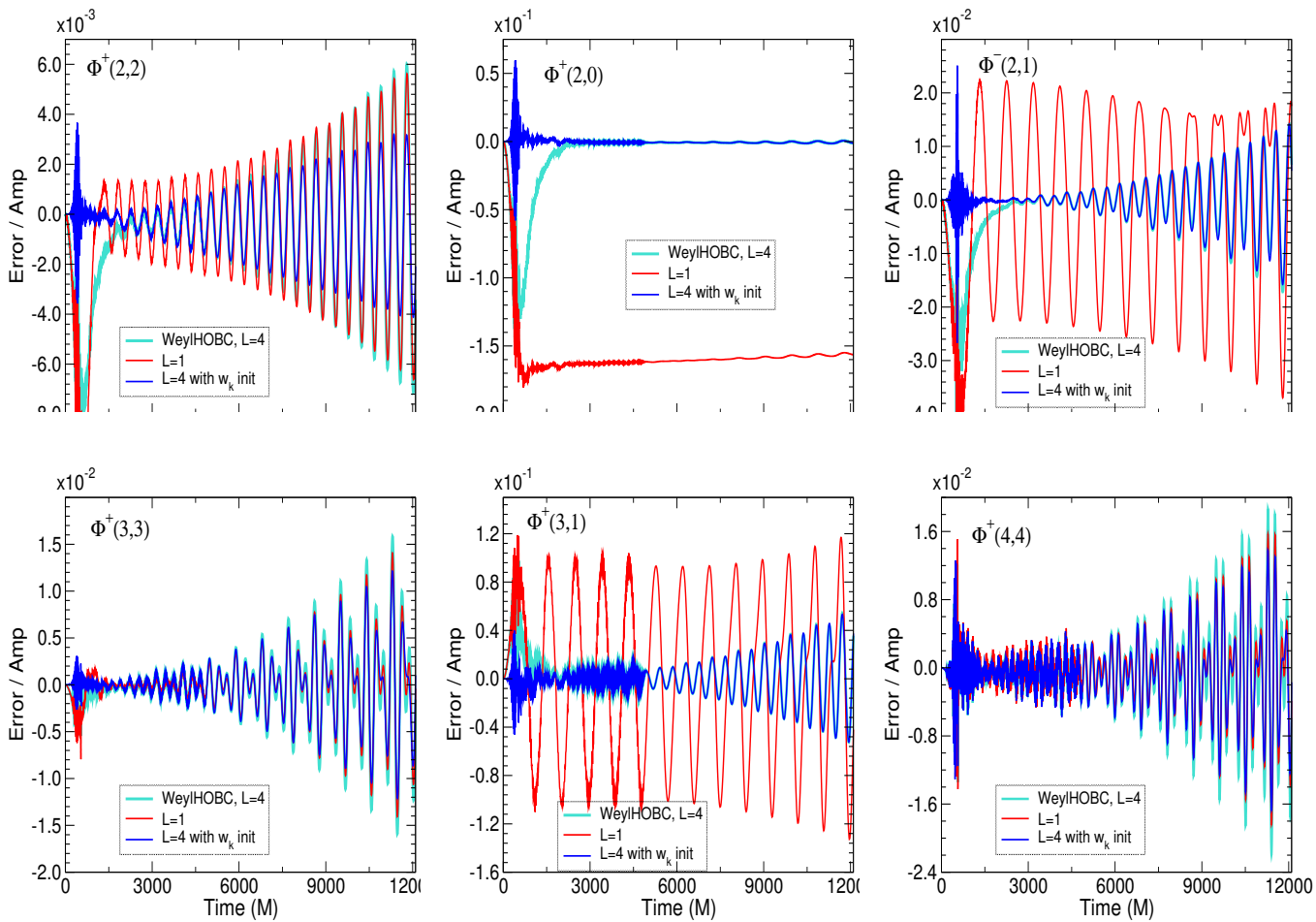

\includegraphics[width=5.9cm, height=6.25cm]{Figure7a.eps}
\includegraphics[width=5.9cm, height=6.25cm]{Figure7b.eps}
\includegraphics[width=5.9cm, height=6.25cm]{Figure7c.eps}
\includegraphics[width=5.9cm, height=6.25cm]{Figure7d.eps}
\includegraphics[width=5.9cm, height=6.25cm]{Figure7e.eps}
\includegraphics[width=5.9cm, height=6.25cm]{Figure7f.eps}
\caption{Same as Figure~\ref{f:bbh-Rbdry250-errors} except: test
  $R_{\rm bdry}=500M$, reference $R_{\rm bdry}=2,646M$, and all waves
  are extracted at $R_{\rm bdry}=500M$.}
\label{f:bbh-Rbdry500-errors}
\end{figure}

We define the error of a mode at time $t$ to be the difference between
$\Phi^{(\pm)}_{\ell m}$ of that mode extracted at $R=R_{\rm bdry}$ and
$\Phi^{(\pm)}_{\ell m}$ of the same mode extracted at the same radius
and time from the reference simulation.  Errors for these six highest
modes from runs with $R_{\rm bdry}=250M$ are shown in
Figure~\ref{f:bbh-Rbdry250-errors}, and those from runs with
$R_{\rm bdry}=500M$ are shown in Figure~\ref{f:bbh-Rbdry500-errors}.
Each error is normalized by the amplitude of the mode
$\Phi^{(\pm)}_{\ell m}$ measured at the extraction radius shortly
after the initial burst of junk radiation has passed. Since the
inspiral for this high mass ratio BBH is so slow, the amplitudes of
the waves remain fairly constant over the time range shown in the
plots.

First, we will discuss the results which are most readily apparent.
For the time range between approximately $1,200M$ and $4,500M$, the
errors in the WeylHOBC runs with $L=4$ and proper
$\text{w}_{k \ell m}$ initialization are about an order of magnitude
smaller than the errors in the $L=1$ runs for the modes
$\Phi^+_{2,2}$, $\Phi^-_{2,1}$, $\Phi^+_{3,3}$, $\Phi^+_{3,1}$, and
$\Phi^+_{4,4}$ in Figure~\ref{f:bbh-Rbdry250-errors} and the modes
$\Phi^+_{2,2}$, $\Phi^-_{2,1}$, and $\Phi^+_{3,1}$ in
Figure~\ref{f:bbh-Rbdry500-errors}. For the time range between
approximately $1,200M$ and $12,000M$, the mode $\Phi^+_{2,0}$ (which
should include the largest GW memory effect) shows about two orders of
magnitude reduction in error in both figures when 4th order WeylHOBC
with proper $\text{w}_{k \ell m}$ initialization is used as opposed to
the freezing-$\Psi_0$ boundary condition. As for the errors in
$\Phi^+_{3,3}$ and $\Phi^+_{4,4}$, they are relatively independent of
boundary condition order (except in Figure~\ref{f:bbh-Rbdry250-errors}
between approximately $650M$ and $4,500M$). We attribute the lack of
improvement in these two modes with HOBCs to the fact that
$\Phi^+_{3,3}$ and $\Phi^+_{4,4}$ are dominated by high-frequency
noise, indicating that they are contaminated by numerical error in the
interior evolution. It is important to note that the normalized errors
in all the modes plotted are small. For example, the normalized errors
in $\Phi^+_{2,2}$ for both $L=1$ and $L=4$ boundary conditions are of
order 1\%. Finally, these figures clearly show that the WeylHOBC
implementation is stable and well-behaved even for a lengthy binary
black hole simulation.  This is not true of the dtHOBC implementation
(not shown), which produces a linear drift superimposed on the actual
waveform.

On further inspection, we see a large initial growth of error at early
times ($t<1,700M$) in both the $L=1$ order runs and the $L=4$ runs in
which the $\text{w}_{k\ell m}^{(\pm)}$ are {\it improperly}
initialized to zero (red and turquoise curves, respectively), for both
$R_{\rm bdry}$ locations.  When $\text{w}_{k\ell m}^{(\pm)}$ are
initialized properly, however, there is no such initial (transient)
growth in error for the $L=4$ order runs.  After this time, both of
the $L=4$ boundary condition runs (with and without proper
$\text{w}_{k\ell m}^{(\pm)}$ initialization) converge together and
track each other fairly closely. This suggests that incorrect
initialization of $\text{w}_{k\ell m}^{(\pm)}$ does not leave a
lasting effect, and therefore one might be willing to suffer this
transient error for simplicity of initialization.

\begin{figure}[!htb]
  \includegraphics[width=8.75cm, height=6.25cm]{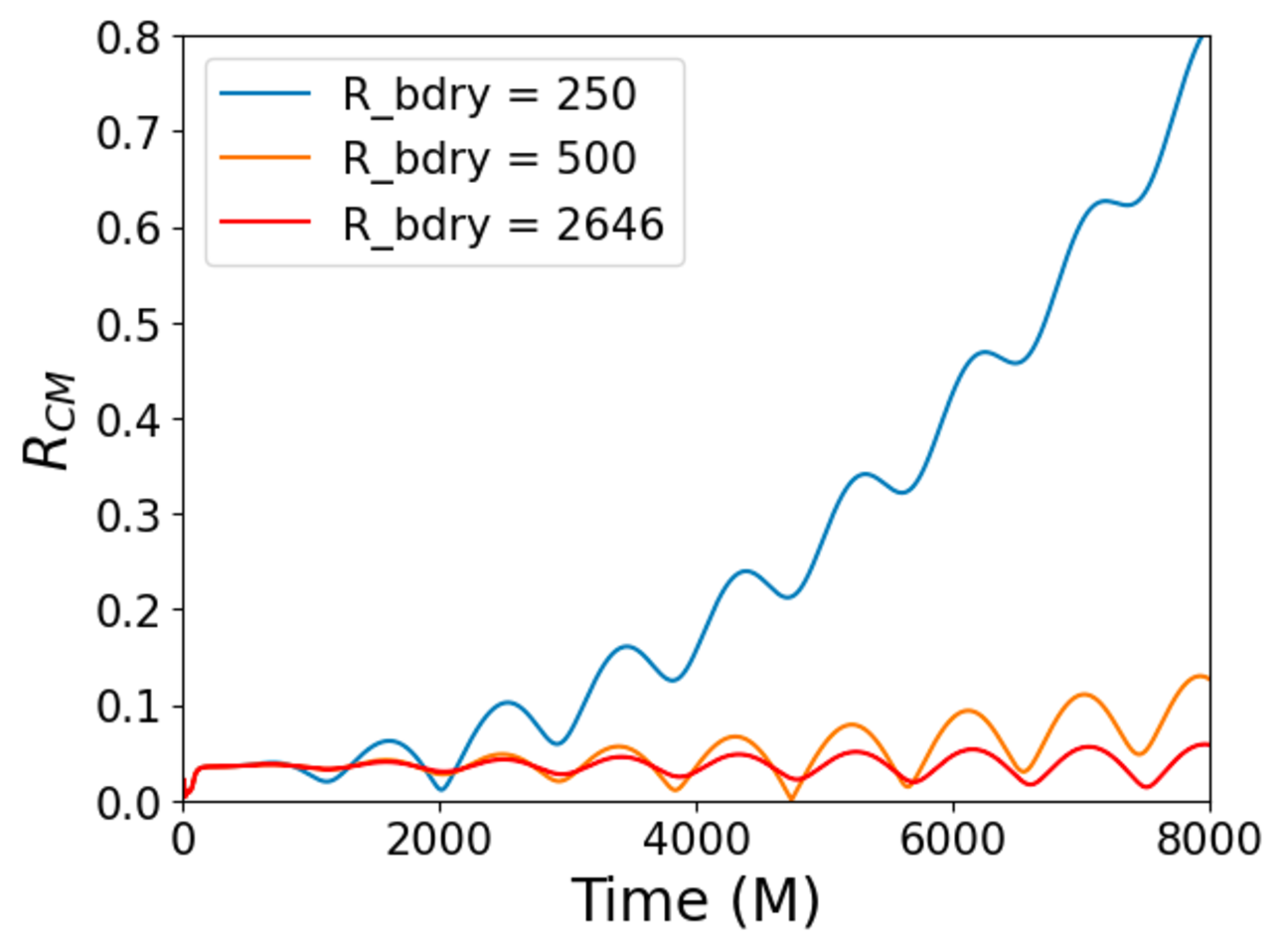}
  \caption{The drift of the coordinate center of mass in BBH
    simulations using the freezing-$\Psi_0$ boundary condition at
    $R_{\rm bdry}=250M$, $500M$, and $2,646M$.}
  \label{f:center-mass-drift}
\end{figure}

Now let us examine in more depth the errors in $\Phi^+_{2,2}$,
$\Phi^-_{2,1}$, $\Phi^+_{3,1}$ and $\Phi^+_{2,0}$ beyond early
times. The errors in the quadrupolar $\Phi^+_{2,2}$ mode for both
boundary condition orders display different behavior depending on
whether the test $R_{\rm bdry}$ is $250M$ or $500M$.  For
$R_{\rm bdry}=250M$, and for $t \lesssim 9,000M$, the error in
$\Phi^+_{2,2}$ is significantly lower when $L=4$ than it is when $L=1$
(see Figure~\ref{f:bbh-Rbdry250-errors}).  Past that time, two effects
are visible.  First, accumulation of error in the $L=4$ simulation
brings the error closer in magnitude to that of the $L=1$ simulation,
which was larger from the beginning but does not grow. Second, all
boundary condition methods show an accelerating drift away from zero.
The latter effect is a known feature of SpEC simulations that
disappears for large $R_{\rm bdry}$ and is associated with a drift in
the coordinate center of mass of the system~\cite{Szil_gyi_2015}.  In
Figure~\ref{f:center-mass-drift}, we plot the drift of the coordinate
center of mass versus time for $R_{\rm bdry}=250M$, $500M$, and
$2,646M$.  We see that the drift is sensitive to outer boundary
location, in agreement with prior studies (see~\cite{Boyle2016}, for
example). The drift grows much more slowly for $R_{\rm bdry}=500M$
than for $R_{\rm bdry}=250M$, and even more slowly for the reference
run which has $R_{\rm bdry}=2,646M$.  Enhancements in numerical
resolution yield a marginal improvement in mitigating this
drift. Altering the order of the boundary conditions, with or without
proper initialization of $\text{w}_{k\ell m}^{(\pm)}$, has no
effect. To understand what causes this drift, first recall that the
HOBCs implemented here are for the {\it physical} characteristic
fields, which represent gravitational wave inflow and outflow.  There
are, in addition, constraint and gauge characteristic fields and
corresponding boundary conditions.  The fact that the drift is
insensitive to the HOBC order suggests that it is not caused by the
physical boundary conditions; in other words, it is not caused by
asymmetric GW reflections producing an unphysical radiation reaction
effect.  The most likely culprit for the coordinate center of mass
drift is the gauge boundary conditions, and indeed, experiments in
progress show that this drift is in fact sensitive to gauge boundary
conditions~\cite{DongzeSun}.
Further examination of the $\Phi^+_{2,2}$ error reveals that the
early-time error for $L=1$ is much lower when the test
$R_{\rm bdry}=500M$ than it is when $R_{\rm bdry}=250M$. Consequently,
the difference in early-time error between $L=1$ and $L=4$ is smaller
when $R_{\rm bdry}=500M$. For the modes $\Phi^-_{2,1}$ and
$\Phi^+_{3,1}$, the fourth order method gives reduced errors for the
duration of the simulation for both boundary locations, even though it
grows after $t \approx 4,500$. This growth in error, which occurs in
most of the tests plotted, could be the result of compounding errors
accumulating from outer boundary reflections and/or gauge BC errors.
\begin{figure}[!htb]
  \includegraphics[width=7.5cm, height=6.25cm]{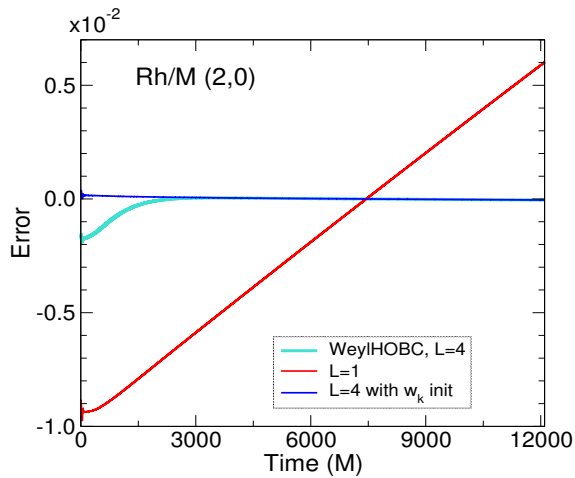}
  \caption{Errors (as measured by the difference between the test and
    reference runs) in the strain ($h$) of the $(2,0)$ mode. The
    strain is computed at future null infinity using the SpECTRE codes
    Cauchy-characteristic evolution (CCE) module. Note that these
    errors are not normalized since the amplitudes from the different
    test run vary. Otherwise, the legend is the same as in
    Figures~\ref{f:bbh-Rbdry250-errors} and
    ~\ref{f:bbh-Rbdry500-errors}. The test $R_{\rm bdry}=500M$ and the
    reference $R_{\rm bdry}=2,646M$.}
  \label{f:cce_errors_20}
\end{figure}
Finally, let us consider the $\Phi^+_{2,0}$ mode more thoroughly. This
mode should contain the largest contribution to GW memory
effects~\cite{Pollney:2010hs,Mitman:2020} and has been difficult to
compute in numerical relativity simulations~\cite{PhysRevD.80.024002}
without Cauchy-characteristic evolution
(CCE)~\cite{Moxon:2020gha,Moxon:2021gbv,Mitman:2020} or
post-processing~\cite{PhysRevD.103.024031} methods. In
Figures~\ref{f:bbh-Rbdry250-errors} and~\ref{f:bbh-Rbdry500-errors},
the $\Phi^+_{2,0}$ mode {\it error} settles to a value offset from
zero with the standard $L=1$ freezing-$\Psi_0$ boundary
condition. With the $L=4$ WeylHOBC method, however, the error in
$\Phi^+_{2,0}$ quickly settles to zero, although if
$\text{w}_{k\ell m}^{(\pm)}$ are incorrectly initialized to zero, the
error has a large initial transient. These results indicate that HOBC
with $L=4$ gives more accurate results for $\Phi^+_{2,0}$ than does
the freezing-$\Psi_0$ boundary condition, with respect to the
reference run.

Nonetheless, studying only the RWZ scalars extracted at finite radii
is not sufficient to understand whether or not HOBCs improve the
resolution of the GW memory in the GW strain, $h$, that is measured at
future null infinity. To verify this, we extract the GW strain at
future null infinity using SpECTRE's CCE
module~\cite{Moxon:2020gha,Moxon:2021gbv,SpECTRE_zenodo}. Furthermore,
we map these waveforms to the superrest frame at 4,000$M$ using a
3-orbit window to ensure that they are in a more reasonable BMS frame
than that which is output by
CCE~\cite{Mitman:2021xkq,PhysRevD.106.084029,Mitman:2024uss}. With
these waveforms, we then study the error in the $(2,0)$ mode, as
measured by the difference between the test and reference runs. This
is highlighted in Fig.~\ref{f:cce_errors_20}, where one sees that the
HOBC curves outperform the standard freezing-$\Psi_{0}$ boundary
condition to a large degree. Furthermore, although not shown in this
figure, comparison to a 3 PN waveform whose parameters match those of
the simulation shows that the net change over time in the $(2,0)$ mode
is much more on par with that of the waveform for the HOBC simulations
than those for the standard freezing-$\Psi_{0}$ boundary condition
simulation. This suggests that the HOBCs indeed improve the resolution
of GW memory in numerical relativity simulations. In the future, we
will perform a more robust comparison of waveforms from simulations
that utilize HOBCs with PN waveforms to illustrate the improved GW
accuracy produced by HOBCs.

The appearance of a small amount of high-frequency noise in the errors
of all modes at early times in Figure~\ref{f:bbh-Rbdry500-errors}
warrants explanation. Notice that this noise disappears rather
abruptly at $t \approx 4,800M$. This high-frequency component is present
only in the reference run and thus shows up in the difference between
waves extracted from the $R_{\rm bdry}=500M$ runs and the reference
run.  Its presence in the reference run is due to high-frequency,
short-wavelength junk radiation in the initial data which pervades the
grid. It persists in the reference run for about the amount of time a
signal takes to pass from near the binary black holes to the reference
outer boundary of $R=2,646M$ and back again to $R_{\rm bdry}=500M$ (which
would be a total time of $4,792M$).  One might have hoped that the
$L=4$ WeylHOBC in the reference run would have eliminated even a
single round of reflections, but this short-wavelength radiation is
not well resolved by the SpEC code at the grid resolutions we have
used, so unphysical backscatter and reflections are possible even with
absorbing boundary conditions.  This high frequency junk radiation
component is not visible in Figure~\ref{f:bbh-Rbdry250-errors} (which
shows the $R_{\rm bdry}=250M$ boundary errors) because these errors
are larger; hence, the scales of the plots are larger.

\begin{figure}[!htb]
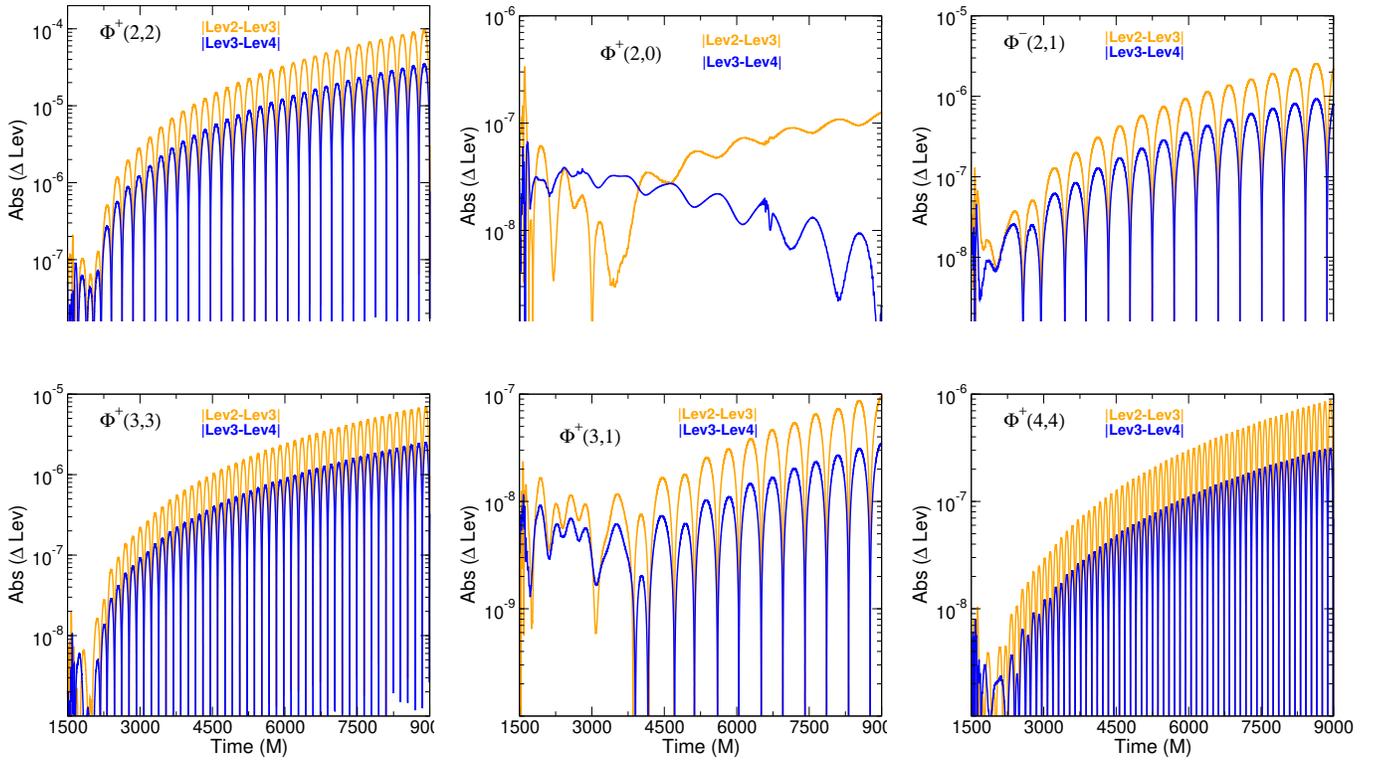

\includegraphics[width=5.9cm, height=5cm]{Figure10a.eps}
\includegraphics[width=5.9cm, height=5cm]{Figure10b.eps}
\includegraphics[width=5.9cm, height=5cm]{Figure10c.eps}
\includegraphics[width=5.9cm, height=5cm]{Figure10d.eps}
\includegraphics[width=5.9cm, height=5cm]{Figure10e.eps}
\includegraphics[width=5.9cm, height=5cm]{Figure10f.eps}
\caption{Convergence plots for BBH simulations using $L=4$ WeylHOBC
  and proper $\text{w}_{k \ell m}$ initialization. Test
  $R_{\rm bdry}=500M$, reference $R_{\rm bdry}=2,646M$, and all waves
  are extracted at $R_{\rm bdry}=500M$. Branching into different
  resolutions (Levs) is done after junk radiation from the initial
  data has left the grid, so all resolutions are identical up to
  $t=1,500M$. Convergence plots for the other cases (runs with $L=4$
  WeylHOBC but no proper $\text{w}_{k \ell m}$ initialization, those
  using the freezing-$\Psi_0$ boundary condition, and reference runs)
  all show similar results.}
\label{f:bbh-Rbdry500-convergence}
\end{figure}

In Figure~\ref{f:bbh-Rbdry500-convergence}, we study the convergence
of our runs with numerical resolution.  We plot the differences
between consecutively higher resolutions for the $R_{\rm bdry}=500M$
runs using 4th order WeylHOBCs with properly initialized
$\text{w}_{k\ell m}^{(\pm)}$. Here ``Lev'' in the figure refers to the
adaptive-mesh-refinement tolerance that determines the grid
resolution; larger Levs have finer grids. All the modes show nice
convergence except for the $\Phi^+_{2,0}$ mode in the time range
$3,000M < t < 4,000M$. We attribute the non-convergence in this time
range to the fact that $\Phi^+_{2,0}$ is primarily non-oscillatory, in
the sense that there is a non-zero offset component to the wave, which
dwarfs the sinusoid, and in addition there is a linear drift
component. Because of these features, when the amplitude of the
$\Phi^+_{2,0}$ mode with one resolution crosses the amplitude of the
same mode with a different resolution at $t \approx 3,000M$, they
linger close to each other. On the other hand, the other modes, which
show nice convergence properties, are more purely oscillatory. Thus,
zero crossings between resolutions are sharp and brief, as seen in the
remaining five panels of Figure~\ref{f:bbh-Rbdry500-convergence}.
\begin{figure}[!htb]
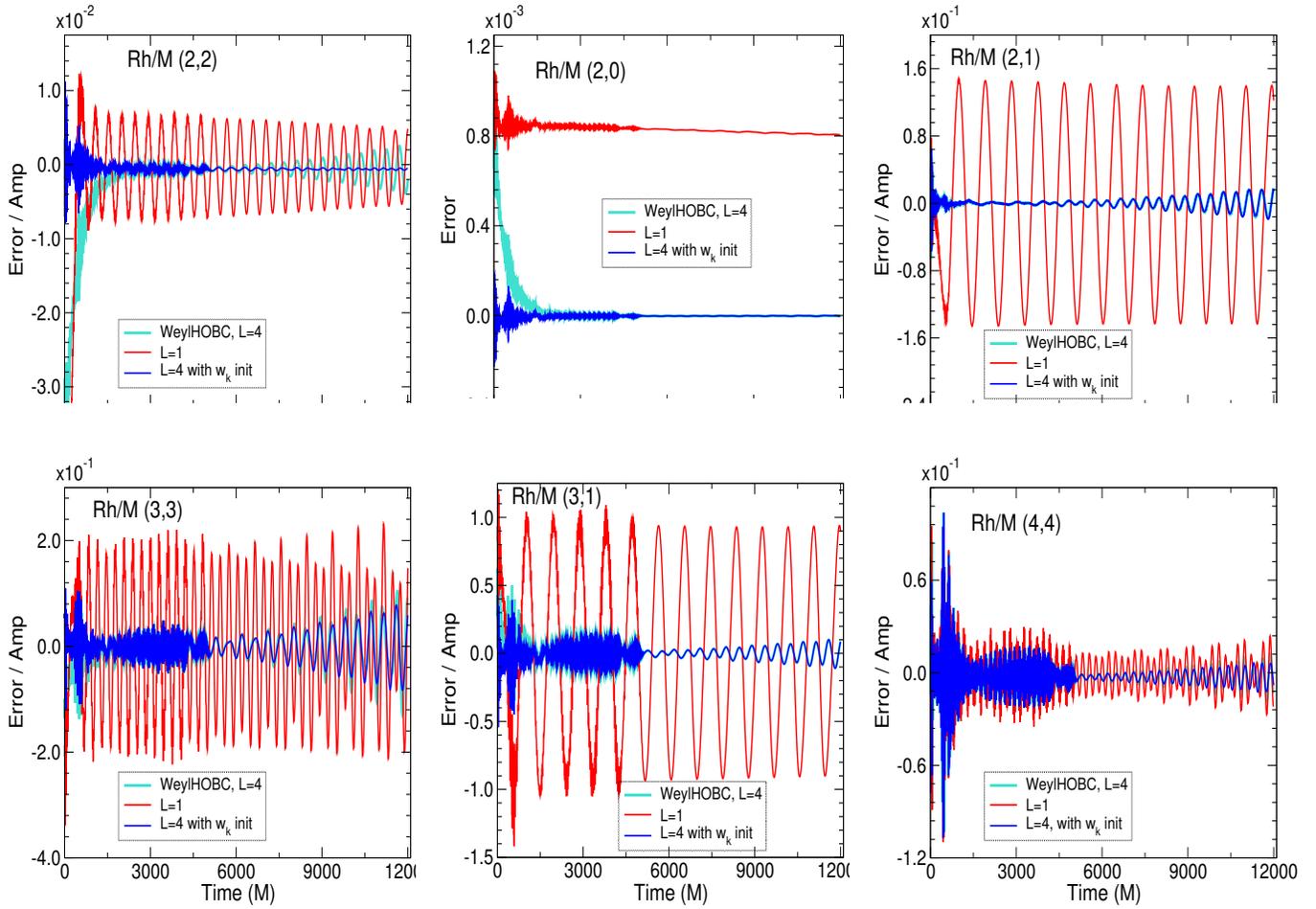

\includegraphics[width=5.9cm, height=6.25cm]{Figure11a.eps}
\includegraphics[width=5.9cm, height=6.25cm]{Figure11b.eps}
\includegraphics[width=5.9cm, height=6.25cm]{Figure11c.eps}
\includegraphics[width=5.9cm, height=6.25cm]{Figure11d.eps}
\includegraphics[width=5.9cm, height=5.9cm]{Figure11e.eps}
\includegraphics[width=5.9cm, height=6.25cm]{Figure11f.eps}
\caption{Errors (as measured by the normalized difference between the
  test and reference runs) in the strain ($h$) which is {\it
    extrapolated} to future null infinity to order N4 using the {\tt
    scri} python code. The legend is the same as in
  Figures~\ref{f:bbh-Rbdry250-errors}, \ref{f:bbh-Rbdry500-errors},
  and~\ref{f:cce_errors_20}. The test $R_{\rm bdry}=500M$ and the
  reference $R_{\rm bdry}=2,646M$.}
\label{f:bbh-strain-extrapolation-errors}
\end{figure}

We conclude our discussion of HOBCs for BBH systems by extrapolating
the gravitational wave strain, $h$, to future null infinity, since
extrapolated strain waveforms are the ultimate product of most
numerical relativity simulations and are used in gravitational wave
detection and interpretation. Our results, shown in
Figure~\ref{f:bbh-strain-extrapolation-errors}, show dramatic
improvement in accuracy with HOBCs for {\it all} the modes
analyzed. In addition, for the $(2,2)$ mode, HOBCs with proper
$\text{w}_{k \ell m}^{(\pm)}$ initialization result in significantly
lower normalized errors than freezing-$\Psi_0$ boundary conditions and
HOBCs without proper initialization, even for times as late as
$12,000 M$. Extrapolation was performed using the {\tt scri} python
code
\cite{mike_boyle_scri_v2022.8.11,Boyle2013,Boyle2014,Boyle2016}. The
strain wave amplitudes $h_+$ and $h_\times$ are defined from the
metric perturbation in the transverse-traceless gauge.  An expression
for $h \equiv h_+ - i h_\times$ in terms of our gauge-invariant
Regge-Wheeler-Zerilli scalars (Eqs.~(\ref{e:Regge-Wheeler}) and
(\ref{e:Zerilli})) is given, to leading order in $1/r$, by
\begin{equation}
  \label{e:WaveAmplitudesFromRWZScalars}
  h = \frac{1}{r}\,\sum_{\ell m}\, \sqrt{\ell(\ell+1)\Lambda}
    \; \left(\Phi^{(+)}_{\ell m} + i \Phi^{(-)}_{\ell m}\right) ~ {}_{-2} Y^{\ell m},
\end{equation}
recalling that $\Lambda \equiv (\ell-1)(\ell+2)$. Note that we follow
the sign conventions outlined in~\cite{Boyle:2019kee} (see Eq.~(10)
and Appendix C). While reading off $h$ directly from the metric
perturbation only works in the transverse-traceless gauge,
Eq.~\ref{e:WaveAmplitudesFromRWZScalars} is valid in any gauge.

There is no extra computational cost when implementing the HOBCs for
BBH simulations. HOBC runs are found to run at about the same speed as
those using {\tt SpEC}'s freezing-$\Psi_0$ boundary
condition. Furthermore, use of HOBC causes no detectable difference in
the violation of the generalized harmonic constraints, neither the
norm of constraint violation over the whole grid nor that of the
outermost spherical domain.

\section{Conclusions and Future Work}
\label{conclusions}

We have presented two new implementations of the high-order absorbing
boundary conditions for the Einstein field equations, WeylHOBC and
dtHOBC, and find WeylHOBC to be clearly superior in accuracy,
robustness, and simplicity of implementation (given a code that
already uses the freezing-$\Psi_0$ boundary condition). We point out
that the problem of coupling a second evolution system to an interior
Cauchy evolution at the latter's boundary occurs in other schemes
besides HOBC, notably in Cauchy characteristic matching.  Our findings
on how best to implement this coupling may be applicable to these
schemes as well. A simple recipe for initializing the
$\text{w}_{k\ell m}^{(\pm)}$ auxiliary functions, which significantly
reduces initial transients, has been demonstrated. This recipe is the
first attempt to satisfy compatibility conditions between the initial
and boundary surfaces. For an unequal mass binary black hole inspiral,
WeylHOBC significantly reduces boundary errors in gravitational
waveforms when compared with the freezing-$\Psi_0$ boundary
condition. It is clear especially in the strain waveforms extrapolated
to future null infinity that these improvements in accuracy occur for
the quadrupolar and subdominant modes, and persist throughout a long
inspiral simulation. We verified that the improvement in accuracy of
the $(2,0)$ mode actually reflects an improvement in the resolution of
the gravitational wave memory by extracting the strain at future null
infinity with Cauchy characteristic extraction, and by comparison with
post-Newtonian waveforms.
    
Although successful implementation of WeylHOBCs for BBHs has been
demonstrated, there are several avenues for future work to gain
increased accuracy in the years ahead. Implementation of the first
order corrections for curvature and backscatter given
in~\cite{Buchman_2007} would improve accuracy even further and allow
for smaller computational grids. By simulating BBHs through merger and
ringdown, one could perhaps reproduce correctly the tail
decay. Implementing WeylHOBCs without demanding the outer boundary
radius to be constant in time would perhaps further increase accuracy
by allowing various outgoing characteristic fields to exit the grid as
the evolution proceeds. Finally, we plan to compare the accuracy of
gravitational waveforms computed with our WeylHOBCs to those computed
with the recent implementation of CCM in the {\tt SpECTRE} code.

The center of mass drift problem seen in SXS long-time BBH simulations
was not improved by our WeylHOBCs, leading one to believe that the
source of this problem is from the gauge boundary conditions. We plan
to investigate whether or not some form of WeylHOBCs could be applied
to the gauge modes and thereby alleviate this problem.

\section{Acknowledgements}

It is a pleasure to thank Olivier Sarbach for providing comments on
the paper prior to publication, for many consultations, and for
hosting L.T.B. at the 2019 BIRS-CMO workshop in Oaxaca,
Mexico. Additionally, we thank Joey Key for her support at UW Bothell,
and Dante Iozzo, Stephen Lau, and Oscar Reula for their insights and
suggestions. Finally, we thank Michael Boyle for the use of his {\tt
  scri} python code. L.T.B. dedicates this paper to the memory of
James M. Bardeen, mentor and friend.

M.D. gratefully acknowledges support from the NSF through grant
PHY-2110287 and from NASA through grant 80NSSC22K0719.
M.M. gratefully acknowledges support from the NSF through grant
AST-2219109 and to the DOE through a Computational Science Graduate
Fellowship. This material is based upon work supported by the
U.S. Department of Energy, Office of Science, Office of Advanced
Scientific Computing Research under Award Number
DE-SC0024386. M.S. gratefully acknowledges support from the Sherman
Fairchild Foundation and from NSF grants PHY-2309211, PHY-2309231, and
OAC-2209656.  T.M.K. and A.E. were supported by NSF Physics REU Award
2050928 and T.M.K. was also supported by NSF CAREER Award
1944412. K.M. was supported by the Sherman Fairchild Foundation and
NSF Grants No. PHY-2011968, PHY-2011961, PHY-2309211, PHY-2309231, and
OAC-2209656 at Caltech.

This report was prepared as an account of work sponsored by an agency
of the United States Government. Neither the United States Government
nor any agency thereof, nor any of their employees, makes any
warranty, express or implied, or assumes any legal liability or
responsibility for the accuracy, completeness, or usefulness of any
information, apparatus, product, or process disclosed, or represents
that its use would not infringe privately owned rights. Reference
herein to any specific commercial product, process, or service by
trade name, trademark, manufacturer, or otherwise does not necessarily
constitute or imply its endorsement, recommendation, or favoring by
the United States Government or any agency thereof. The views and
opinions of authors expressed herein do not necessarily state or
reflect those of the United States Government or any agency thereof.

\bibstyle{prd}
\bibliography{References}
\end{document}